\documentclass[aps,10pt,prx,twocolumn,showpacs,amsmath,amssymb,superscriptaddress,floatfix]{revtex4-2}

\usepackage[english]{babel}
\usepackage{blindtext}
\usepackage{latexsym}
\usepackage{amssymb}
\usepackage{physics}
\usepackage{amsmath}
\usepackage{bm}
\usepackage{amsfonts}
\usepackage{relsize}
\usepackage{xcolor}
\usepackage{verbatim}
\usepackage{bbold}
\usepackage{slashed}
\usepackage{appendix}
\usepackage{graphicx}
\usepackage[normalem]{ulem}
\usepackage[colorlinks = true,
            linkcolor = blue,
            urlcolor  = blue,
            citecolor = blue,
            anchorcolor = blue]{hyperref}
\usepackage{graphicx} 
\usepackage{dcolumn} 

\newcommand{\msml}[1]{{\mathsmaller{\mathsmaller{\mathsmaller{#1}}}}}
\newcommand{\smsqr}{\msml{\square}}
\newcommand{\bs}[1]{\mathbf{#1}}

\newcommand{\kk}{\bs{k}}

\newcommand{\qb}{\bs{q}}
\newcommand{\QQ}{\bs{Q}}

\newcommand{\RR}{\bs{R}}

\renewcommand{\bar}{\overline}

\newcommand{\up}{\uparrow}
\newcommand{\down}{\downarrow}
\newcommand{\Neel}{N\'eel}

\definecolor{ao(english)}{rgb}{0.0, 0.5, 0.0}

\begin{document}

\title{Spin stiffnesses and stability of magnetic order \\ in the lightly doped two-dimensional Hubbard model}

\author{Demetrio Vilardi}
\affiliation{Max Planck Institute for Solid State Research, D-70569 Stuttgart, Germany}

\author{Pietro M.~Bonetti}
\affiliation{Department of Physics, Harvard University, Cambridge, Massachusetts 02138, USA}
\affiliation{Max Planck Institute for Solid State Research, D-70569 Stuttgart, Germany}

\author{Walter Metzner}
\affiliation{Max Planck Institute for Solid State Research, D-70569 Stuttgart, Germany}

\date{\today}

\begin{abstract}
We analyze the density dependence of the spin stiffnesses and the stability of magnetic order with respect to quantum fluctuations in the two-dimensional Hubbard model close to half-filling. The stiffnesses are computed from the spin susceptibility obtained from a random phase approximation in a magnetically ordered state. For a sizable next-to-nearest neighbor hopping amplitude and a moderate Hubbard interaction, the mean-field ground state is a N\'eel antiferromagnet in the electron doped regime at and above half-filling, and a planar circular spiral state in the hole doped regime below half-filling. Upon electron doping, the N\'eel stiffness decreases smoothly and not very steeply. By contrast, the in-plane and out-of-plane stiffnesses in the spiral state drop abruptly at half-filling. The out-of-plane stiffness even drops to zero, and then increases again very slowly upon increasing hole doping.
At finite temperatures, the N\'eel-to-spiral transition is shifted into the hole doped regime, the stiffnesses are continuous functions of the density, and they vanish at the transition.
For small hole doping, the spin stiffnesses describe the quantum fluctuations only in a small momentum range, which shrinks to zero upon approaching half-filling.
Using the above results, we show that the quantum ground state of the lightly electron doped Hubbard model remains N\'eel ordered, while quantum fluctuations probably destroy the spiral long-range order in the hole doped regime, giving rise to a quantum disordered state with a spin gap.
\end{abstract}


\maketitle


\section{Introduction}

The two-dimensional Hubbard model has attracted much interest as a minimal model capturing the most salient properties of the valence electrons moving in the copper-oxide planes of cuprate high-temperature superconductors \cite{Anderson1987, Zhang1988}.
Inspite of remarkable progress in the development of quantum many-body methods over the past decades, only small fractions of the phase diagram of the 2D Hubbard model have been firmly established \cite{Arovas2022, Qin2022}. In case of pure nearest neighbor hopping, the ground state is doubtlessly a N\'eel ordered antiferromagnet at half-filling \cite{Qin2016}, for any interaction strength, and a collinear spin-charge stripe state at the special densities $n = 1 \pm \frac{1}{8}$ for sufficiently strong interactions \cite{Qin2020}. In presence of hopping amplitudes beyond nearest neighbors, the N\'eel order at half-filling sets in only above a certain critical interaction strength. At sufficiently weak coupling the ground state of the 2D Hubbard model is a d-wave superconductor in a broad density range around half-filling \cite{Neumayr2003, Raghu2010, Metzner2012}.

Besides the special cases just mentioned, there are no conclusive results on magnetic order in the ground state of the 2D Hubbard model away from half-filling. Exact numerical methods are limited to relatively small finite size lattices. Monte Carlo methods are severely limited by the notorious sign problem. From approximate methods, such as Hartree-Fock theory, dynamical mean-field theory and its cluster extensions, and the functional renormalization group, essentially three types of magnetic order emerged as plausible candidates: N\'eel order \cite{Lin1987, Hofstetter1998, Langmann2007, Chubukov1992, Igoshev2010, Yamase2016}, planar circular spiral order \cite{Shraiman1989, Jayaprakash1989, Dombre1990, Fresard1991, Shraiman1992, Chubukov1992, Chubukov1995, Kotov2004, Igoshev2010, Igoshev2015, Yamase2016, Bonetti2020, Goremykin2023}, and collinear spin-charge stripe order \cite{Schulz1989, Zaanen1989, Machida1989, Poilblanc1989, Schulz1990, Kato1990, Seibold1998, Fleck2000, Fleck2001, Raczkowski2010, Timirgazin2012, Peters2014, Zheng2017}.
In a recent comprehensive Hartree-Fock study of the two-dimensional Hubbard model at a moderate interaction strength these three states indeed covered almost the entire magnetic phase diagram \cite{Scholle2023}. A more complex state with non-collinear planar magnetic order was found in a narrow density regime between spiral and stripe order \cite{Scholle2024}. Messy states in an intermediate regime between spiral order and a magnetic phase with domain walls have been observed previously in a Hartree-Fock study of a three-band Hubbard model \cite{Chiciak2018}.
Away from half-filling, the magnetic order may coexist with superconductivity \cite{Reiss2007, Wang2014, Yamase2016, Lichtenstein2000, Capone2006, Aichhorn2006, Kancharla2008, Zheng2016, Xu2024}.

At finite temperatures, magnetic order is rigorously excluded by the Mermin-Wagner theorem \cite{Mermin1966}: the SU(2) spin symmetry of the Hubbard model cannot be spontaneously broken in two dimensions. In the ground state, instead, quantum fluctuations of the order parameter reduce the magnetic order found in mean-field theories, but not necessarily to zero. The impact of low energy quantum fluctuations can be analyzed within a non-linear sigma model for the transverse fluctuations of the order parameter \cite{Auerbach1994}. The key parameters in this model are the spin stiffnesses, which parametrize the energy of long wavelength fluctuations. In a spiral state one needs to distinguish spin stiffnesses related to in-plane and out-of-plane fluctuations \cite{Bonetti2022}.
The non-linear sigma model maintains a magnetically ordered state for sufficiently large stiffnesses, and yields a quantum disordered state for small stiffnesses.
Mean-field spin stiffnesses also play an important role in SU(2) gauge theories of the pseudogap phase in cuprates at finite temperatures \cite{Bonetti2022gauge, Goremykin2024, Vasiliou2024, Bultinck2024}.

In this article we compute the spin stiffnesses for the lightly doped Hubbard model with a moderate interaction strength within mean-field theory. We include a sizable next-to-nearest neighbor hopping amplitude, which, in the context of cuprates, is more realistic than the Hubbard model with pure nearest neighbor hopping. The mean-field ground state then exhibits N\'eel order in a density range at and above half-filling, and spiral order below half-filling. While the stiffness evolves continuously as a function of density on the electron doped side ($n >1$), there are discontinuous drops at half-filling to strongly reduced values upon hole doping ($n < 1$). The spatial out-of-plane stiffness of the spiral state even vanishes upon approaching half-filling from below. We derive a number of analytic results for the wave vector of the spiral state at low hole doping, and for the size of the stiffness drops. We also present and analyze results for the stiffnesses at finite temperatures.
Finally, we apply our results for the stiffnesses to address the issue of stability of the magnetically ordered ground state with respect to quantum fluctuations of the order parameter. From the well established fact of N\'eel order at half-filling \cite{Qin2016}, and the continuity of the stiffnesses for $n \geq 1$, we can firmly conclude that N\'eel order persists in a finite density interval above half-filling. By contrast, the drastic drop especially of the spatial out-of-plane stiffness suggests the possibility of a discontinuous transition to a quantum disordered ground state with a spin gap upon hole doping.

The remainder of the paper is structured as follows.
In Sec.~\ref{sec: II} we compute the spin stiffnesses for spiral magnetic order and discuss their density dependence. Based on these results, in Sec.~\ref{sec: III} we analyze the stability of magnetic order near half-filling with respect to quantum fluctuations. A summary and outlook in Sec.~\ref{sec: IV} concludes the presentation.


\section{Spiral order and spin stiffnesses} \label{sec: II}

In this section we analyze the spin stiffnesses of spiral magnetic order in the two-dimensional Hubbard model within mean-field theory.

\subsection{Hubbard model}

The Hubbard Hamiltonian for spin-$\frac{1}{2}$ lattice fermions with a local repulsive interaction $U>0$ reads \cite{Arovas2022, Qin2022}
\begin{equation} \label{eq: HubbardHamiltonian}
 H = \sum_{j,j',s} t_{jj'} c^\dagger_{j,s} c^{\phantom\dagger}_{j',s} +
 U \sum_j n_{j,\up} n_{j,\down} \, ,
\end{equation}
where $c_{j,s}$ ($c^\dagger_{j,s}$) annihilates (creates) a fermion with spin orientation $s \in \{\uparrow,\downarrow\}$ on the lattice site $j$, and $n_{j,s} = c^\dagger_{j,s} c^{\phantom\dagger}_{j,s}$.
The hopping matrix $t_{jj'}$ depends only on the distance between the sites $j$ and $j'$.
We consider the case of a square lattice with nearest and next-to-nearest hopping amplitudes, that is, $t_{jj'} = -t$ if $j$ and $j'$ are nearest neighbor sites, $t_{jj'} = -t'$ if $j$ and $j'$ are next-to-nearest neighbors, and $t_{jj'} = 0$ otherwise.
Fourier transforming, this hopping matrix yields the dispersion
\begin{equation} \label{eq: dispersion}
 \epsilon_\kk = -2t (\cos k_x + \cos k_y) - 4t' \cos k_x \cos k_y \, .
\end{equation}
We use the nearest neighbor hopping amplitude $t$ as our energy unit.


\subsection{Mean-field theory}

Recently, an unbiased mean-field phase diagram of the two-dimensional Hubbard model at moderate interaction strengths has been established \cite{Scholle2023, Scholle2024}. For pure nearest neighbor hopping, the mean-field ground state is a N\'eel state only at half-filling, which turns into a spin-charge stripe state upon doping. For a finite next-to-nearest neighbor hopping amplitude $t'<0$, the N\'eel state remains stable upon electron doping. Upon hole-doping instead, the N\'eel state survives only for weak interaction strengths, and is replaced by a planar spiral state for moderate interactions. A stripe state is favorable only at sufficiently large doping if $t' \neq 0$. A spiral state is found for the hole-doped Hubbard model also at strong coupling in a dynamical mean-field approximation \cite{Bonetti2020, Goremykin2023}.
The mean-field formalism for the planar spiral magnetic order is adopted from established literature~\cite{Penn1966,Chubukov1992,Kampf1996}. 

For planar spiral magnetic order all spin expectation values $\langle \bs{S}_j \rangle$ lie in the same (arbitrary) plane and have the same amplitude $m = |\langle \bs{S}_j \rangle|$ for all sites $j$. We choose the spins to be aligned in the $xy$ plane. The spiral order then has the form
\begin{equation} \label{eq: spiral}
 \langle \bs{S}_{j} \rangle = m \left[
 \cos(\bs{Q}\cdot \bs{R}_j) \, \bs{e}_x + \sin(\bs{Q}\cdot \bs{R}_j) \, \bs{e}_y
 \right] \, ,
\end{equation}
where $\bs{Q}$ is a fixed wave vector, $\bs{R}_j$ is the position vector of the lattice site $j$, and $\bs{e}_\alpha$ are unit vectors with directions $\alpha = x,y,z$. 
The components of the spin operator $\bs{S}_j$ can be expressed in terms of fermion creation and annihilation operators as
\begin{equation} \label{eq: Sj}
 S^a_j = \frac{1}{2}\sum_{s,s'} c^\dagger_{j,s} \sigma^a_{ss'} c_{j,s'} \, ,
\end{equation}
where $\sigma^a$ with $a=1,2,3$ are the Pauli matrices.
In the following it will be convenient to include also the charge operator $\rho_j$ as $S^0_j = \frac{1}{2} \rho_j$ with the 2x2 identity matrix $\sigma^0$ in Eq.~\eqref{eq: Sj}.
Fourier transforming, the Eqs.~\eqref{eq: spiral} and \eqref{eq: Sj} yield the relation
\begin{equation}
 m = \int_{\bs{k}} \langle
 c_{\bs{k},\up}^\dagger c_{\bs{k}+\bs{Q},\down}^{\phantom\dagger} \rangle \, ,
\end{equation}
where $c_{\bs{k},s}^{\phantom\dagger}$ and $c_{\bs{k},s}^\dagger$ are creation and annihilation operators in momentum space, and $\int_{\bs{k}} = \int \frac{d^2\bs{k}}{(2\pi)^2}$.
For spiral order of the form \eqref{eq: spiral}, the mean-field Hamiltonian can be written in momentum representation as~\cite{Kampf1996}
\begin{equation} \label{eq: Ham spiral}
H_{\rm MF} = \int_\kk \sum_{s} \xi_\kk c_{\kk,s}^\dagger c_{\kk,s}^{\phantom\dagger} +
\Delta \int_\kk c_{\kk,\up}^\dagger c_{\kk+\QQ,\down}^{\phantom\dagger} + \mathrm{h.c.} \, ,
\end{equation}
where $\xi_\kk = \epsilon_\kk - \mu$, and $\Delta = Um$ is the energy gap associated with the spiral order.
Diagonalizing $H_{\rm MF}$ yields two quasiparticle bands
\begin{equation}
 E_\kk^\pm = \frac{1}{2} (\xi_\kk + \xi_{\kk+\QQ}) \pm
 \sqrt{\frac{1}{4}(\xi_\kk - \xi_{\kk+\QQ})^2 + \Delta^2} \, .
\end{equation}
The gap is determined by the self-consistency equation~\cite{Kampf1996}
\begin{equation} \label{eq: gap equation}
 \Delta = U \int_\kk \frac{\Delta}{2e_\kk}\left[ f(E^-_\kk) - f(E^+_\kk) \right] \, ,
\end{equation}
where $e_\kk = \sqrt{h_\kk^2 + \Delta^2}$ with $h_\kk = \frac{1}{2} (\xi_\kk - \xi_{\kk+\QQ})$,
and $f(x) = (e^{x/T} + 1)^{-1}$ is the Fermi function.
The optimal wave vector $\QQ$ is obtained by minimizing the mean-field free energy~\cite{Penn1966,Kampf1996}
\begin{equation}
 F(\QQ) = - T \int_\kk \sum_{\ell} \ln\left( 1 + e^{-E_\kk^\ell/T} \right)
 + \frac{\Delta^2}{U} + \mu n \, ,
\end{equation}
at fixed density $n$.

It is useful to introduce rotated fermion operators \cite{Kampf1996}
\begin{equation} \label{eq: spinrot} 
\begin{split}
 \tilde{c}_{j} &=
 e^{-\frac{i}{2} \QQ \cdot \RR_j} e^{\frac{i}{2} \QQ \cdot \RR_j \sigma^3} c_j \, , \\
 \tilde{c}^\dagger_{j} &= c^\dagger_j
 e^{-\frac{i}{2} \QQ \cdot \RR_j \sigma^3} e^{\frac{i}{2} \QQ \cdot \RR_j } \, ,
\end{split}
\end{equation}
where $c_j = (c_{j,\up},c_{j,\down})$ and $c_j^\dagger =
(c_{j,\up}^\dagger,c_{j,\down}^\dagger)$.
In the new basis $\tilde{c}_{j} = (\tilde{c}_{j,\up},\tilde{c}_{j,\down})$, the spins appear ordered ferromagnetically in the x-direction.
In momentum space, the transformation \eqref{eq: spinrot} corresponds to
$\tilde{c}_\kk = (\tilde{c}_{\kk,\up},\tilde{c}_{\kk,\down}) =
(c_{\kk,\up},c_{\kk+\QQ,\down})$.
In the rotated basis, the inverse Green function has the form
\begin{equation}
 \widetilde{G}^{-1}(\kk,i\nu) = \left( \begin{matrix}
 i\nu - \xi_\kk & -\Delta \\ -\Delta & i\nu  - \xi_{\kk+\QQ}
 \end{matrix} \right) \, .
\end{equation}
Using a partial fraction decomposition, the Green function can also be written as \cite{Bonetti2022}
\begin{equation} \label{eq: G decomp}
 \widetilde{G}(\kk,i\nu) =
 \frac{1}{2} \sum_{\ell=\pm} \frac{u_\kk^\ell}{i\nu - E_\kk^\ell} \, ,
\end{equation}
with the matrix coefficients
\begin{equation}
 u_\kk^\ell = \sigma^0 + \ell \frac{h_\kk}{e_\kk} \sigma^3 +
 \ell \frac{\Delta}{e_\kk} \sigma^1 \, .
\end{equation}
%


\subsection{Spin-charge susceptibility}

The imaginary time spin-charge susceptibility can be defined as~\cite{Kampf1996,Bonetti2022}
\begin{equation} \label{eq: chi}
 \chi^{ab}_{jj'}(\tau) = \langle \mathcal{T} S_j^a(\tau) S_{j'}^b(0) \rangle \, ,
\end{equation}
where $S_j^a(\tau)$ with $a \in \{0,1,2,3\}$ is the spin/charge operator from Eq.~\eqref{eq: Sj} in the imaginary time Heisenberg representation, and $\mathcal{T}$ represents the time-ordering operator.

In a spiral state, translation invariance is broken, and this property is shared by the susceptibility. The susceptibility in the rotated spin basis,
\begin{equation} \label{eq: chitilde}
 \tilde\chi^{ab}_{jj'}(\tau) =
 \langle \mathcal{T} \tilde{S}^a_j(\tau) \tilde{S}^b_{j'}(0) \rangle \, ,
\end{equation}
with $\tilde S_j^a = \frac{1}{2} \tilde{c}_j^\dagger \sigma^a \tilde{c}_j^{\phantom\dagger}$, remains instead translation invariant.
One can transform the susceptibility to the rotated basis and back by using the formula
\begin{equation} \label{eq: chi rot}
 \chi_{jj'}(\tau) = \mathcal{M}_j \tilde\chi_{jj'}(\tau) \mathcal{M}^{T}_{j'} \, ,
\end{equation}
where $\chi_{jj'}(\tau)$ and $\tilde\chi_{jj'}(\tau)$ are $4\times4$ matrices with matrix elements $\chi_{jj'}^{ab}(\tau)$ and $\tilde\chi_{jj'}^{ab}(\tau)$, respectively.
The right hand side of Eq.~\eqref{eq: chi rot} is a matrix product with the rotation matrix
\begin{equation} \label{eq: chi M chi M}
 \mathcal{M}_j = \left( \begin{array}{cccc}
 1 & 0 & 0 & 0 \\
 0 & \cos (\QQ \cdot \RR_j) & -\sin (\QQ \cdot \RR_j) & 0 \\
 0 & \sin (\QQ \cdot \RR_j) & \cos (\QQ \cdot \RR_j) & 0 \\
 0 & 0 & 0 & 1
 \end{array} \right) \, .
\end{equation}

We denote the Fourier transform of $\chi_{jj'}(\tau)$ by $\chi(\qb,\qb',i\omega_n)$, and its analytic continuation to real frequencies $i\omega_n \to \omega + i0^+$ by $\chi(\qb,\qb',\omega)$. Due to the broken translation invariance, the susceptibility has non-zero components not only for $\qb' = \qb$, but also for $\qb' = \qb \pm \QQ$ and $\qb' = \qb \pm 2\QQ$.
Fourier transforming Eq.~\eqref{eq: chi rot}, one obtains the following relations for the momentum and spin diagonal components of the susceptibility \cite{Kampf1996, Bonetti2022}
\begin{eqnarray} \label{eq: trafo dia}
 \chi^{00}(\qb,\omega) &=& \tilde\chi^{00}(\qb,\omega) ,  \\
 \chi^{11}(\qb,\omega) &=& \chi^{22}(\qb,\omega) \label{eq: chi22 unr} \\
 &=& \frac{1}{4} \big[ \tilde\chi^{11}(\qb\!+\!\QQ,\omega) +
 \tilde\chi^{11}(\qb\!-\!\QQ,\omega) \nonumber \\
 &&+ \tilde\chi^{22}(\qb\!+\!\QQ,\omega) + \tilde\chi^{22}(\qb\!-\!\QQ,\omega) \nonumber \\
 &&+ 2i \tilde\chi^{12}(\qb\!+\!\QQ,\omega) + 2i\tilde\chi^{21}(\qb\!-\!\QQ,\omega) \big] ,
 \hskip 5mm \nonumber \\
 \chi^{33}(\qb,\omega) &=& \tilde\chi^{33}(\qb,\omega) \, .
\end{eqnarray}
%
%
%

In random phase approximation (RPA), the susceptibility in the rotated spin frame is given by~\cite{Kampf1996, Bonetti2022}
\begin{equation} \label{eq: rpa}
 \tilde\chi(\qb,\omega) = \tilde\chi_0(\qb,\omega)
 \left[ {\rm I}_4 - \Gamma_0 \tilde\chi_0(\qb,\omega) \right]^{-1} \, ,
\end{equation}
where $\Gamma_0 = 2U {\rm diag}(-1,1,1,1)$ is the bare interaction vertex, $\rm I_4$ is the $4\times4$ identity matrix, and $\tilde\chi_0(\qb,\omega)$ is composed of the matrix elements
\begin{eqnarray}
 \tilde\chi_0^{ab}(\qb,\omega) &=&
 - \frac{1}{4} \int_\kk T \sum_{\nu_m} \Tr \big[ \sigma^a \widetilde{G}(\kk,i\nu_m)
 \nonumber \\
 && \times \sigma^b \widetilde{G}(\kk+\qb,i\nu_m+i\omega_n) \big]
 \Big|_{i\omega_n \rightarrow \omega + i0^+} \hskip 3mm
\end{eqnarray}
is the bare susceptibility.
Using the decomposition \eqref{eq: G decomp}, the Matsubara frequency sum can be carried out, yielding
\begin{equation} \label{eq: chi0t}
 \tilde \chi_0^{ab}(\qb,\omega) =
 -\frac{1}{8} \sum_{\ell,\ell'=\pm} \int_\kk \mathcal{A}^{ab}_{\ell\ell'}(\kk,\qb)
 \frac{f(E_\kk^\ell) - f(E_{\kk+\qb}^{\ell'})}
 {\omega + i0^+ + E_\kk^\ell - E_{\kk+\qb}^{\ell'}} \, ,
\end{equation}
with the coefficients
\begin{equation}
 \mathcal{A}^{ab}_{\ell\ell'}(\kk,\qb) =
 \frac{1}{2} \Tr \left( \sigma^a u_\kk^\ell \sigma^b u_{\kk+\qb}^{\ell'} \right) \, .
\end{equation}
%


\subsection{Goldstone poles}

The diagonal components of the spin susceptibility $\chi^{aa}(\qb,\omega)$ with $a=1,2,3$ have poles at $\qb = \pm\QQ$ and $\omega =0$, which are associated with the Goldstone modes. In the spiral state there are three Goldstone modes, one corresponding to \emph{in-plane} fluctuations, and two corresponding to \emph{out-of-plane} fluctuations \cite{Rastelli1985, Chandra1990, Shraiman1992, Kampf1996}. For our choice of spin order in the $xy$ plane, the in-plane and out-of-plane modes are reflected by poles in $\chi^{11} = \chi^{22}$ and $\chi^{33}$, respectively. In the rotated spin basis, where all spins are aligned along the $x$ direction, the in-plane mode appears as a (single) pole in $\tilde{\chi}^{22}(\qb,\omega)$ at $(\qb,\omega) = (\bs{0},0)$, while $\tilde{\chi}^{11}(\qb,\omega)$ remains bounded at low energies.

Near the poles, the spin susceptibilities in the rotated frame have the form
\begin{eqnarray}
 \tilde\chi^{22}(\qb,\omega) &\simeq&
 \frac{m^2}{J_{\alpha\beta}^\Box q_\alpha q_\beta - Z^\Box \omega^2} \, ,
\label{eq: Goldstone22} \\
 \tilde\chi^{33}(\qb,\omega) &\simeq &
 \frac{m^2}{J_{\alpha\beta}^\perp(q_\alpha \mp Q_\alpha)(q_\beta \mp Q_\beta)
 - Z^\perp \omega^2} \, , \hskip 5mm
\label{eq: Goldstone33}
\end{eqnarray}
with Einstein's summation convention for repeated Greek indices.
We have not written (imaginary) Landau damping terms in the denominator, which are of the same (quadratic) order as the other terms in the in-plane susceptibility, and of higher (cubic) order in the out-of-plane susceptibility \cite{Bonetti2022}.
The coefficients in the above expansion are the \emph{spatial} and \emph{temporal spin stiffnesses}, which are related to second derivatives of the inverse susceptibilities at the Goldstone poles as \cite{Bonetti2022,factor2}
\begin{eqnarray}
\label{eq: Jin def}
 J^{\Box}_{\alpha\beta} &=&
 \frac{1}{2} m^2 \, \partial^2_{q_\alpha q_\beta} \! \left. \left[
 \tilde{\chi}^{22}(\qb,0) \right]^{-1} \right|_{\qb=\bs{0}} \, , \\
\label{eq: Jout def}
 J^{\perp}_{\alpha\beta} &=&
 \frac{1}{2} m^2 \, \partial^2_{q_\alpha q_\beta} \! \left. \left[
 \tilde{\chi}^{33}(\qb,0) \right]^{-1} \right|_{\qb=\pm\QQ} \, ,
\end{eqnarray}
and
\begin{eqnarray}
\label{eq: Zin def}
 Z^{\Box} &=&
 - \frac{1}{2} m^2 \, \partial^2_\omega \! \left. \left[
 \tilde{\chi}^{22}(\bs{0},\omega) \right]^{-1} \right|_{\omega=0} \, , \\
\label{eq: Zout def}
 Z^{\perp} &=&
 - \frac{1}{2} m^2 \, \partial^2_\omega \! \left. \left[
 \tilde{\chi}^{33}(\pm\QQ,\omega) \right]^{-1} \right|_{\omega=0} \, .
\end{eqnarray}
%


\subsection{Special case: N\'eel order}

N\'eel order can be viewed as a special case of spiral order with $\QQ = (\pi,\pi)$.
All spins are collinear in the N\'eel state, and the SU(2) symmetry is only partially broken: a U(1) symmetry remains. Hence, there are only two linearly independent Goldstone modes, corresponding to fluctuations perpendicular to the spin alignment in the N\'eel state. The symmetry of the N\'eel state with respect to rotations around the axis of spin alignment (the $x$ axis in our representation) implies that these two Goldstone modes are degenerate, and are thus described by the same spatial and temporal stiffnesses, $J_{\alpha\beta} = J \delta_{\alpha\beta}$ and $Z$, respectively.
This degeneracy can also be obtained from the above formulae for the stiffnesses in the spiral state, by using the symmetry relation $\tilde{\chi}^{22}(\omega,\qb) = \tilde{\chi}^{33}(\omega,\qb+\QQ)$, which is specific for the N\'eel state \cite{Bonetti2022}.

Since $2\QQ = (2\pi,2\pi) \equiv \bs{0}$ in the N\'eel state, the off-diagonal components of the susceptibilities with $\qb' = \qb \pm 2\QQ$ contribute actually to the diagonal components, yielding the simpler relations \cite{Bonetti2022}
\begin{equation}
 \chi^{aa}(\qb,\omega) = \tilde{\chi}^{aa}(\qb+\QQ,\omega) \, ,
\end{equation}
for $a = 1,2$.


\subsection{Spatial stiffnesses}

The evaluation of the RPA susceptibilities generally involves a $4\times4$ matrix inversion. However, reflection and other symmetries facilitate the evaluation of the RPA stiffnesses, so that relatively simple expressions can be derived \cite{Bonetti2022, Bonetti2022ward}.

\subsubsection{In-plane stiffness}

For the spatial in-plane stiffness one obtains \cite{Bonetti2022}
{\setlength{\abovedisplayskip}{3pt} 
 \setlength{\belowdisplayskip}{3pt} 
\begin{equation} \label{eq: Jin}
\begin{split}
 J^{\Box}_{\alpha\beta} &=
 - 2\Delta^2 \Big[ \partial^2_{q_\alpha q_\beta} \tilde{\chi}^{22}_0(\qb,0) \\
 &+ 2\sum_{ab=0,1} \partial_{q_\alpha}\tilde{\chi}^{2a}_0(\qb,0) \bar\Gamma^{ab}(\qb,0) \partial_{q_\beta}\tilde{\chi}^{b2}_0(\bs{q},0)\Big] \Big|_{\qb=\bs{0}} \, ,
\end{split}
\end{equation}
} 
where the matrix $\bar\Gamma(q)$ represents the RPA effective interaction in the subspace spanned by the charge channel and the spin channel in $x$-direction,
{\setlength{\abovedisplayskip}{3pt} 
 \setlength{\belowdisplayskip}{3pt} 
\begin{eqnarray}
 \bar\Gamma(q) &=&
 \left[ {\rm I}_2 -
 \left( \begin{array}{cc} -2U & 0 \\ 0 & 2U \end{array} \right)
 \left( \begin{array}{cc} \tilde{\chi}_0^{00}(q) & \tilde{\chi}_0^{01}(q) \\
 \tilde{\chi}_0^{10}(q) & \tilde{\chi}_0^{11}(q) \end{array} \right)
 \right]^{-1} \nonumber \\
 &\times& \left( \begin{array}{cc} -2U & 0 \\ 0 & 2U \end{array} \right) \, ,
\end{eqnarray}
}
with $q = (\qb,\omega)$. The matrix elements $\bar\Gamma^{ab}(q)$ with $a,b \in \{0,1\}$ are all finite for $\omega = 0$ and $\qb \to \bs{0}$.

The momentum derivatives on the right hand side of Eq.~\eqref{eq: Jin} can be evaluated by applying them to the bare susceptibilities as given in Eq.~\eqref{eq: chi0t}.
For the derivative in the first term one obtains \cite{Bonetti2022ward}
{\setlength{\abovedisplayskip}{3pt} 
 \setlength{\belowdisplayskip}{3pt} 
\begin{align} \label{eq: Jin0}
 & \left. \partial^2_{q_\alpha q_\beta}
 \tilde{\chi}^{22}_0(\bs{q},0) \right|_{\qb=0} = \nonumber \\
 & \int_\kk \gamma^\alpha_{\kk}\gamma^\beta_{\kk+\QQ} \!
 \left[ \frac{f(E_\kk^-) - f(E_\kk^+)}{8e^3_\kk} +
 \frac{f'(E_\kk^+) + f'(E_\kk^-)}{8e^2_\kk} \right] ,
\end{align}
}
where $f'(x) = df/dx$, and $\gamma_\kk^\alpha = \partial_{k_\alpha} \epsilon_\kk$. The momentum derivatives in the second term can be written as
\begin{align}
 & \left. \partial_{q_\alpha} \tilde{\chi}_0^{20}(\qb,0) \right|_{\qb=0} =
 - \left. \partial_{q_\alpha} \tilde{\chi}_0^{02}(\qb,0) \right|_{\qb=0} = \nonumber \\[1mm]
 & i \Delta \int_\kk \partial_{k_\alpha} h_\kk
 \left[ \frac{f(E_\kk^-) - f(E_\kk^+)}{8e^3_\kk} +
 \frac{f'(E_\kk^+) + f'(E_\kk^-)}{8e^2_\kk} \right] \, ,
\end{align}
and
\begin{align}
 & \left. \partial_{q_\alpha} \tilde{\chi}_0^{21}(\qb,0) \right|_{\qb=0} =
 - \left. \partial_{q_\alpha} \tilde{\chi}_0^{12}(\qb,0) \right|_{\qb=0} \nonumber \\[2mm]
 & = i \int_\kk h_\kk \partial_{k_\alpha} g_\kk \,
 \frac{f(E_\kk^-) - f(E_\kk^+)}{8e^3_\kk} \nonumber \\
 & + i \int_\kk \sum_\ell \left[
 h_\kk \partial_{k_\alpha} g_\kk + \ell e_\kk \partial_{k_\alpha} h_\kk \right]
 \frac{f'(E_\kk^\ell)}{8e_\kk^2} \, ,
\end{align}
where $g_\kk = \frac{1}{2} \left( \xi_\kk + \xi_{\kk+\QQ} \right)$.


\subsubsection{Out-of-plane stiffness}

For the spatial out-of-plane stiffness one obtains the simple expression \cite{Bonetti2022}
\begin{equation}
 J^{\perp}_{\alpha\beta} = \left. - 2 \Delta^2 \partial^2_{q_\alpha q_\beta}
 \tilde{\chi}^{33}_0(\qb,0) \right|_{\qb=\QQ} \, .
\end{equation}
It is fully determined by the bare susceptibility.
Applying the momentum derivatives to $\tilde{\chi}^{33}_0(\qb,0)$ as given by Eq.~\eqref{eq: chi0t} yields the expression \cite{Bonetti2022ward, factor2}
\begin{widetext}
\begin{align} \label{eq: Jout}
 J^{\perp}_{\alpha\beta} &=
 \frac{1}{4}\int_\kk \sum_{\ell,\ell'=\pm} \left( 1 - \ell \frac{h_\kk}{e_\kk} \right)
 \left( 1 + \ell'\frac{h_{\kk+\QQ}}{e_{\kk+\QQ}} \right)
 \gamma^\alpha_{\kk+\QQ}\gamma^\beta_{\kk+\QQ}
 \frac{f(E^\ell_\kk) - f(E^{\ell'}_{\kk+\QQ})}{E^\ell_\kk - E^{\ell'}_{\kk+\QQ}}
 \nonumber \\
 & -\frac{1}{4}\int_\kk \sum_{\ell=\pm}\left[ \left( 1 - \ell\frac{h_\kk}{e_\kk}\right)^2
 \gamma^\alpha_{\kk+\QQ} + \frac{\Delta^2}{e^2_\kk}\gamma^\alpha_\kk \right]
 \gamma^\beta_{\kk+\QQ}f'(E^\ell_\kk)
 - \frac{1}{4}\int_\kk \sum_{\ell=\pm} \left[\frac{\Delta^2}{e^2_\kk}
 (\gamma^\alpha_{\kk+\QQ}-\gamma^\alpha_\kk) \right] \gamma^\beta_{\kk+\QQ}
 \frac{f(E^\ell_\kk) - f(E^{-\ell}_\kk)}{E^\ell_\kk - E^{-\ell}_\kk} \, .
\end{align}
\end{widetext}
%


\subsection{Numerical results: ground state}

\begin{figure}[tb]
 \centering
 \includegraphics[width=0.9\linewidth]{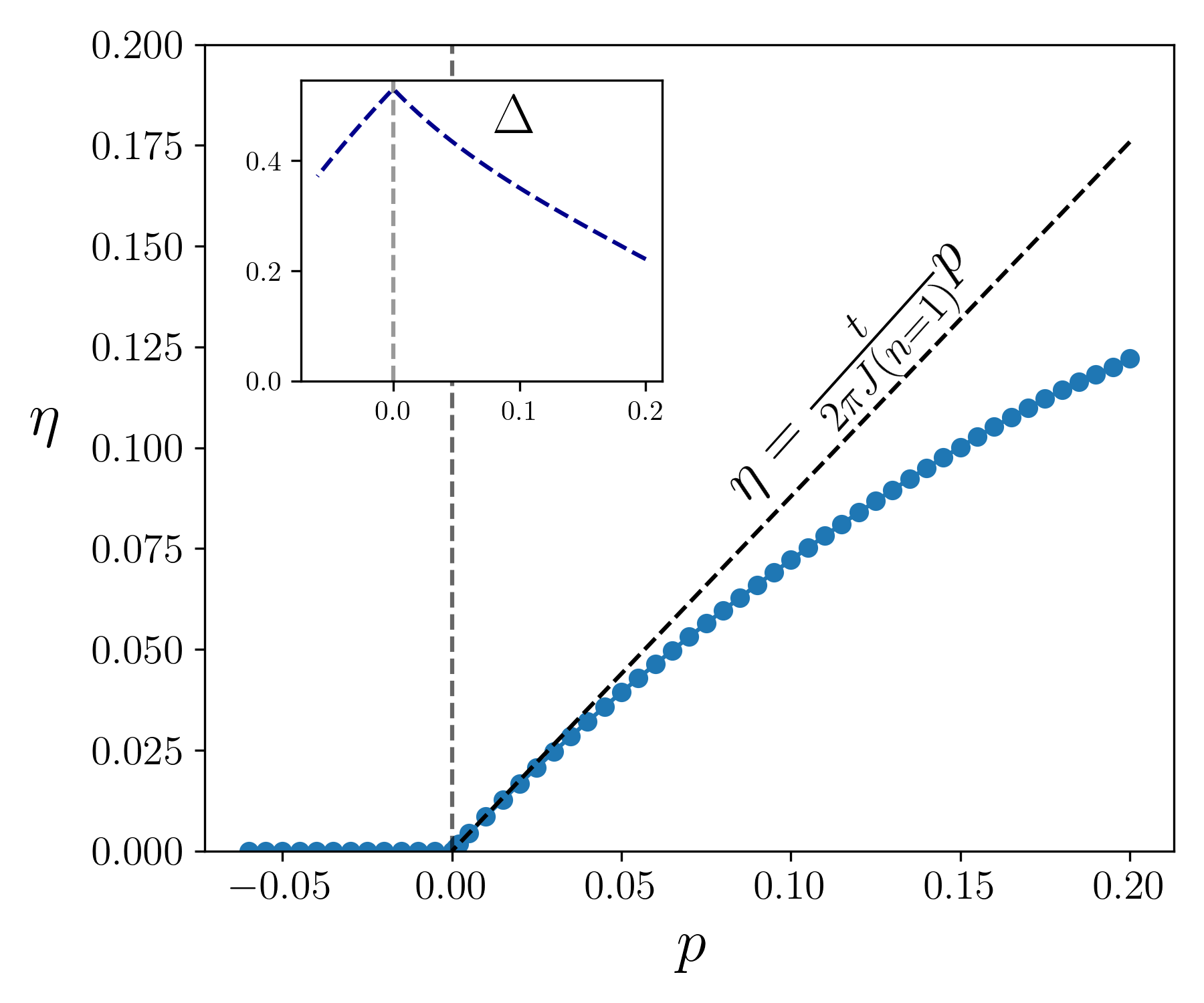}
 \caption{Incommensurability $\eta$ as a function of doping $p = 1-n$ in the ground state.
 The dashed line represents the asymptotically linear behavior for small $p$.
 Inset: Magnetic gap $\Delta$ as a function of doping.}
 \label{fig: eta}
\end{figure}
\begin{figure}[tb]
 \centering
 \includegraphics[width=0.9\linewidth]{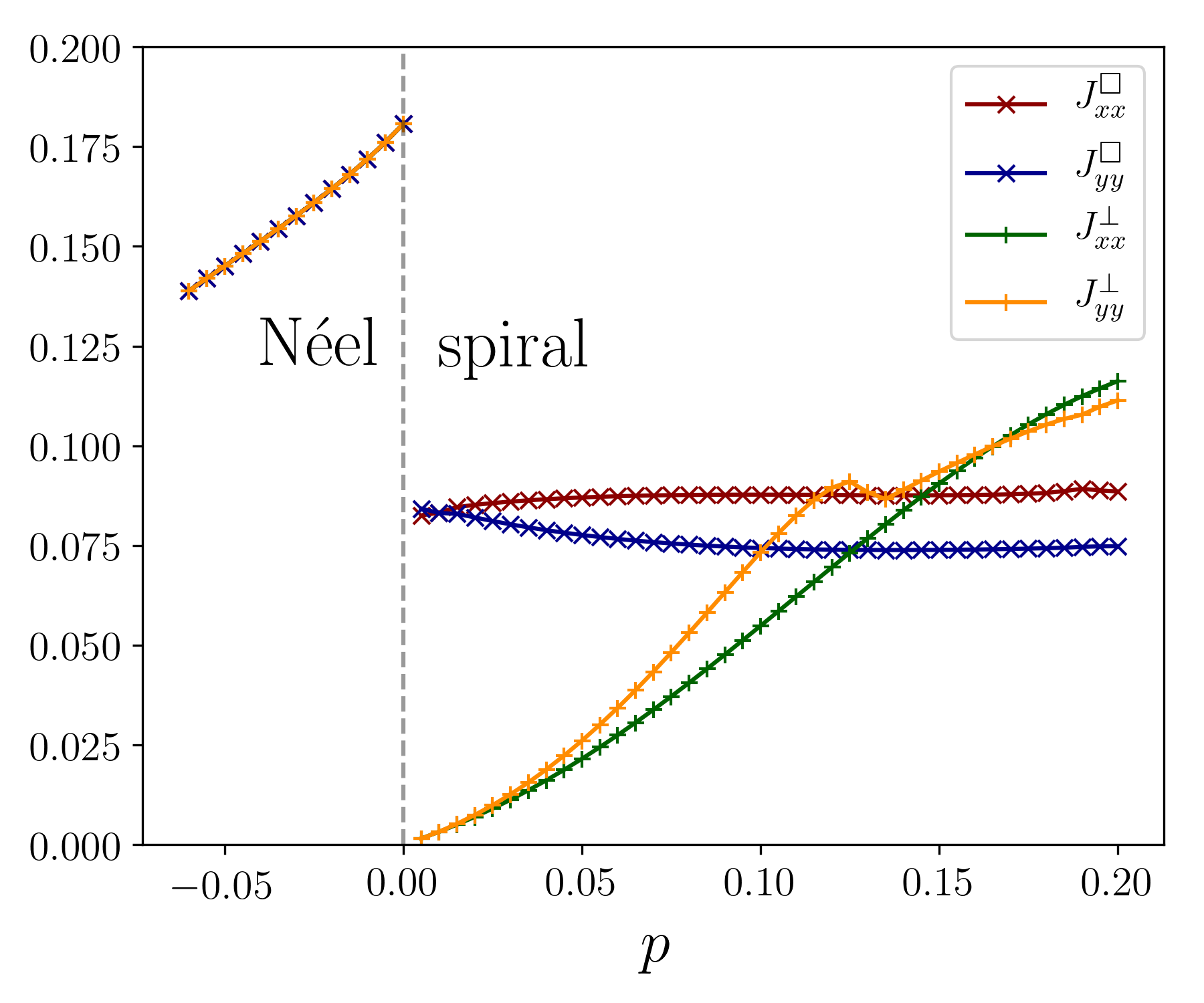}
 \caption{Spatial stiffnesses as a function of doping $p$ in the ground state.
 For the spiral state in the hole-doped regime, there are distinct in-plane and
 out-of-plane stiffnesses, with distinct $xx$ and $yy$ components. In the N\'eel
 state, for $p \leq 0$, both Goldstone modes are characterized by the same
 isotropic stiffness.}
 \label{fig: J_groundstate}
\end{figure}

In this section we show some numerical results for the order parameter and the spatial spin stiffnesses in the ground state as a function of doping $p=1-n$.
We choose $t=1$ as energy unit, $t'=-0.2t$, and a Hubbard interaction $U=2.35t$.
Viewed as an effective interaction obtained from a renormalization group flow \cite{Wang2014, Yamase2016, Vilardi2020}, this interaction strength corresponds to a bare Hubbard interaction near $U=4t$.
For these parameters, the ground state is N\'eel ordered at half-filling and in the electron doped regime ($n \geq 1$), while in the weakly hole doped regime spiral order with a wave vector of the form $\QQ = (\pi-2\pi\eta,\pi)$ with $\eta > 0$ is obtained. There is a two-fold degeneracy in that the incommensurability $\eta$ could of course appear equivalently in the second component of $\QQ$, instead of the first one.

Within mean-field theory, there is a sizable range of moderate coupling strengths $U$ exhibiting qualitatively the same phase diagram near half-filling \cite{Igoshev2010}. The results presented in this section, and the underlying physics of spiral magnetic order being favored over N\'eel order upon small finite hole doping, are robust for any value of $U$ for which the N\'eel order is unstable towards spiral order at any small finite doping $p > 0$. For the next-nearest-neighbor hopping $t'=-0.2t$ considered here, this range of (effective) interaction $U$ is approximately $2.22t \lesssim U \lesssim 3t$~\cite{Igoshev2010}.

For weaker couplings, $2.1t < U < 2.22t$, the N\'eel state remains the ground state at small finite doping before transitioning to other phases. Furthermore, for $U \lesssim 2.1t$, the system becomes entirely paramagnetic near half-filling due to the finite value of $t'$, thus preventing the formation of any ordered magnetic state. At stronger couplings, the mean-field approximation loses its reliability, requiring the use of more sophisticated techniques, such as dynamical mean-field theory~\cite{Georges1996}. Therefore, the choice of $U = 2.35t$ is not a rigid restriction but a representative value that exhibits the N\'eel to spiral instability upon small hole doping, justifying the use of this value to showcase the general behavior of the model.

At larger hole-doping eventually a spin-charge stripe state has a lower energy \cite{Scholle2023, Scholle2024}. Although the focus of this work is on the regime of weak hole doping, where a spiral state has the lowest energy, we show results for spiral states also at larger hole doping, that is, in a regime where the ground state is actually stripe ordered. This is to see trends in the doping dependence in a wider regime.

In Fig.~\ref{fig: eta} we show the incommensurability $\eta$ and, in the inset, the magnetic gap $\Delta$ as a function of doping.
The incommensurability $\eta$ decreases linearly while approaching half-filling, and vanishes continuously at $n=1$. There is a simple relation between $\eta$ and doping $p$ in the low doping limit, as we will show below.

In Fig.~\ref{fig: J_groundstate} the stiffnesses $J^{\Box}$ and $J^{\perp}$ are plotted as a function of the doping. As already observed in Ref.~\cite{Bonetti2022,Bonetti2022gauge}, both stiffnesses are discontinuous at half-filling. As we will see below, one can analytically compute the size of the jumps and analyze their origin.
Remarkably, the out-of-plane stiffness $J^{\perp}$ tends to zero for $p \to 0^+$, which we will also prove analytically.
This feature was not observed in Refs.~\cite{Bonetti2022,Bonetti2022gauge}, due to increasing numerical inaccuracies for smaller hole doping.

\begin{figure}[tb]
 \centering
 \includegraphics[width=0.9\linewidth]{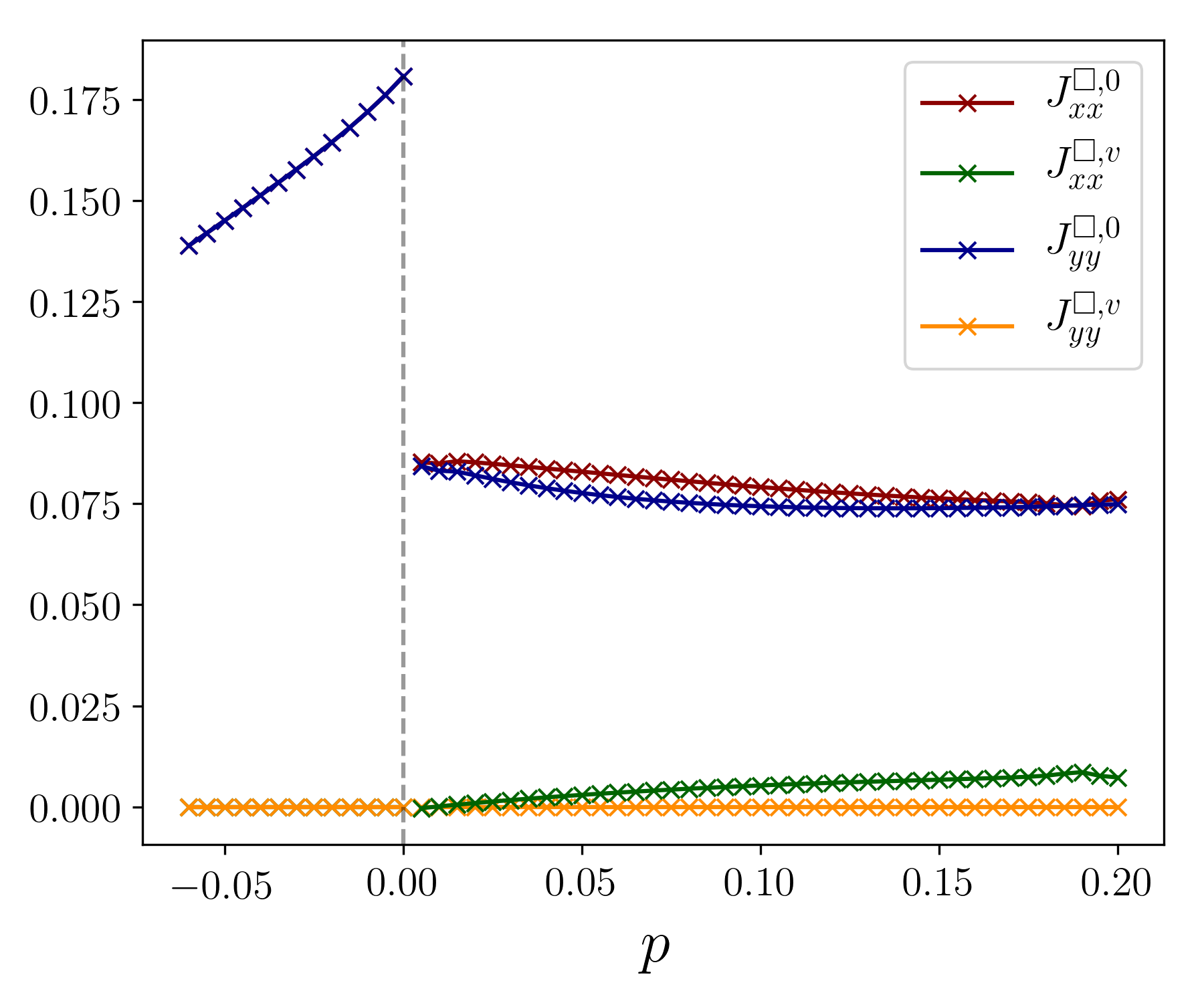}
 \caption{Bare contributions $J_{\alpha\alpha}^{\Box,0}$ and vertex corrections
 $J_{\alpha\alpha}^{\Box,v}$ to the spatial in-plane stiffness
 $J_{\alpha\alpha}^\Box$ as a function of doping. There are no vertex corrections
 in the N\'eel state, and in the spiral state only $J_{xx}^{\Box,v}$ is non-zero.}
 \label{fig: Jin vertex}
\end{figure}
Contrary to the out-in-plane stiffness, which is fully determined by a single particle-hole bubble, the in-plane stiffness has in general vertex corrections, given by the second term in Eq.~\eqref{eq: Jin}. In Fig.~\ref{fig: Jin vertex} we compare the bare component and its vertex correction.
At half-filling and for electron doping, the vertex corrections vanish since the functions
$\partial_{q_\alpha} \tilde \chi^{2a}(\qb,0)$ and
$\partial_{q_\alpha} \tilde \chi^{b2}(\qb,0)$ are identically zero in the N\'eel state.
In the spiral state with $\QQ = (\pi-2\pi\eta,\pi)$, which corresponds to N\'eel type order in the $y$ direction, the vertex correction $J_{yy}^{\Box,v}$ remains zero, too.
The only remaining vertex correction $J_{xx}^{\Box,v}$ is generally small compared to its bare counterpart, and vanishes at small doping.

\begin{figure}[tb]
 \centering
 \includegraphics[width=0.8\linewidth]{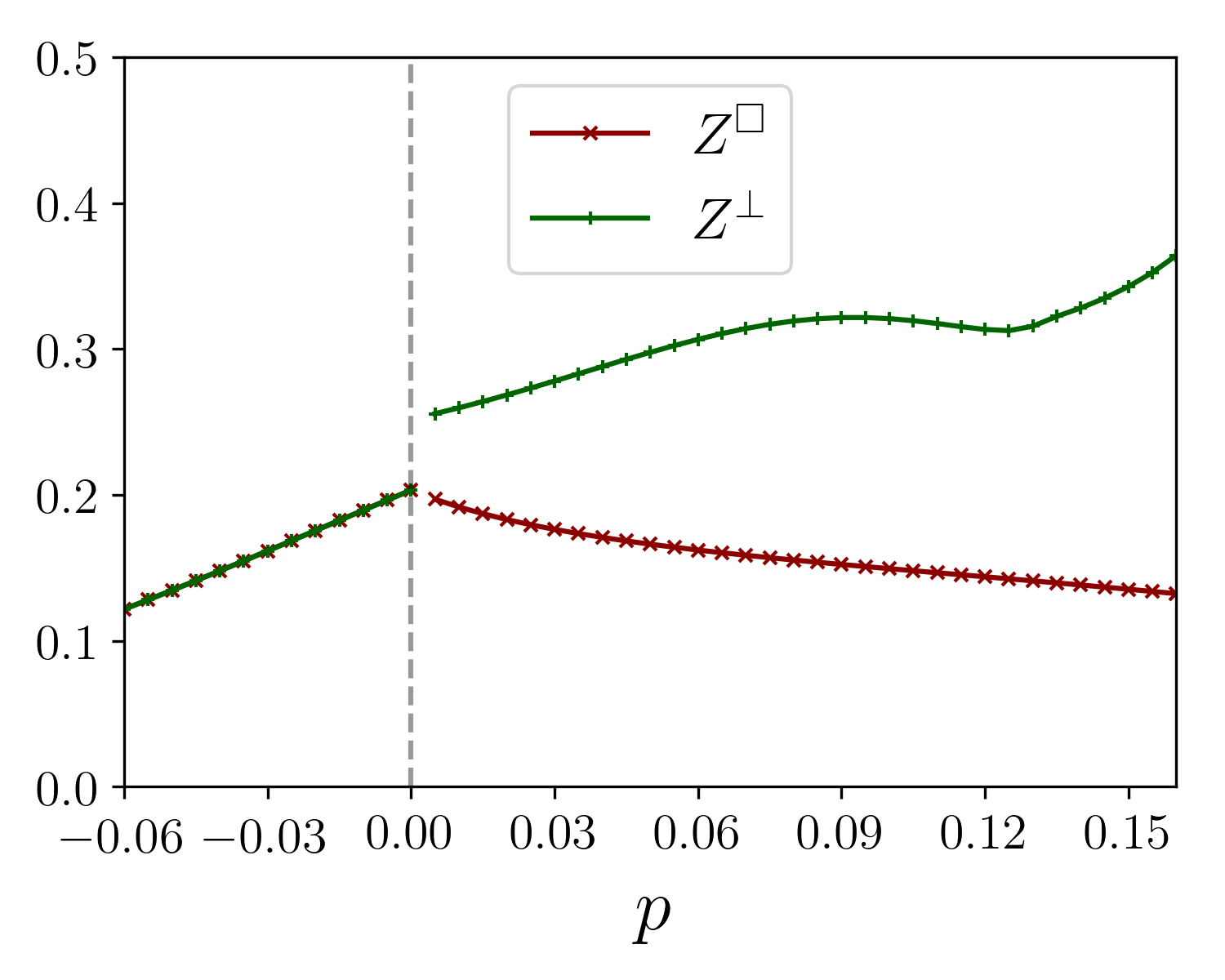}
 \caption{Temporal stiffnesses as a function of doping $p$ in the ground state.
 For the spiral state in the hole-doped regime, there are distinct in-plane and
 out-of-plane stiffnesses.}
 \label{fig: Za}
\end{figure}
In Fig.~\ref{fig: Za} we show the density dependence of the temporal stiffnesses. In the hole doped spiral regime there are again distinct in-plane and out-of-plane stiffnesses, $Z^\Box$ and $Z^\perp$, respectively, see Eqs.~\eqref{eq: Zin def} and \eqref{eq: Zout def}. Explicit expressions can be found in Refs.~\cite{Bonetti2022, Bonetti2022ward}. In the electron doped N\'eel regime, both Goldstone modes have the same temporal stiffness $Z$. The in-plane stiffness is continuous, while the out-of-plane stiffness exhibits a pronounced drop at half-filling.


\subsection{Low doping expansion} \label{sec: low doping}

In this section we evaluate the spatial spin stiffnesses of the spiral ground state in the limit of low hole doping, following partially Ref.~\cite{Chubukov1995}. In particular, we will derive a simple formula for the discontinuities of the stiffnesses at half-filling, and we will show that the out-of-plane stiffness vanishes upon approaching half-filling from the hole-doped regime.


\subsubsection{Hole pockets}

\begin{figure}[t]
    \centering
    \includegraphics[width=0.98\linewidth]{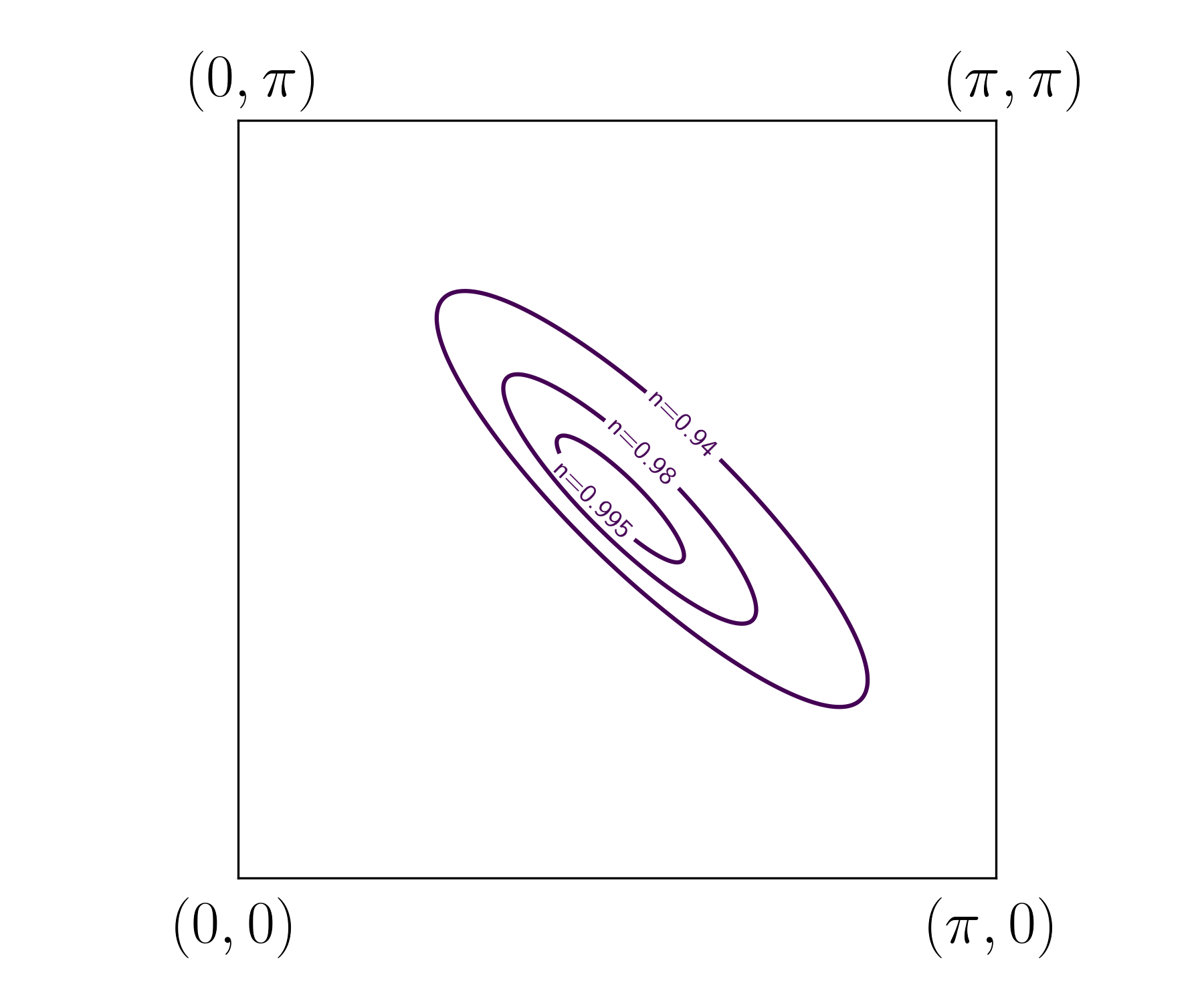}
    \caption{Hole pockets in the first quarter of the Brillouin zone, for electron
    densities $n < 1$ close to half-filling. }
    \label{fig: pockets}
\end{figure}
At half-filling the lower quasiparticle band $E_\kk^-$ is completely filled, while the upper band $E_\kk^+$ remains empty.
For low hole-doping, small hole pockets are formed in the lower quasiparticle band, as shown in Fig.~\ref{fig: pockets}, with a Fermi surface defined by $E^-_\kk = 0$.

The centers of the pockets are defined by the maxima of $E^-_\kk = 0$ in momentum space, where $\nabla_\kk E^-_\kk = 0$. For $\QQ = (\pi - 2\pi\eta,\pi)$, their positions are given by the four points
\begin{equation}
 \kk_{ss'} = \left( s \frac{\pi}{2} + \pi\eta, s' \frac{\pi}{2} \right) \, ,
\end{equation}
with $s,s' \in \{+,-\}$.
They also satisfy the independent equations $\epsilon_\kk = \epsilon_{\kk+\QQ}$ and
$\nabla_\kk (\epsilon_\kk + \epsilon_{\kk+\QQ}) = 0$.
Note that these points depend only on the incommensurability $\eta$, not on the details of the dispersion $\epsilon_\kk$, as long as the hopping amplitudes beyond nearest neighbors remain sufficiently small compared to $t$.
In the N\'eel state all four pockets are equal in size. For $\eta > 0$, the two pockets on the right, centered around $(\pi/2 + \pi\eta, \pm \pi/2)$, are larger than the pockets around $(-\pi/2 + \pi\eta, \pm \pi/2)$ on the left.

In the low doping limit, we can expand $E_\kk^-$ around the maxima at $\kk_{ss'}$ to quadratic order~\cite{Chubukov1995}
\begin{equation} \label{eq: disp low doping}
 E_\kk^- = E_s^-
 - \frac{1}{2} \tilde\kk^T \cdot M_s^{-1} \tilde\kk + {\cal O}(|\tilde\kk|^3) \, ,
\end{equation}
with $\tilde\kk = \kk - \kk_{ss'}$.
For a dispersion of the form \eqref{eq: dispersion}, one finds
$E_s^- = - \Delta - \mu + 2st \sin(\pi\eta)$, and
\begin{align}
 & M_s^{-1} = \\
 & \left(
 \begin{array}{cc}
 \frac{2 t [t \cos (2\delta) + s t \Delta \sin\delta]}{\Delta } &
 s \frac{4 \left[t^2 - 2stt' \sin\delta + t'\Delta \right] \cos\delta}{\Delta} \\
 s \frac{4 \left[t^2 - 2stt'\sin\delta + t'\Delta \right] \cos\delta}{\Delta} &
 \frac{4 [t - 2st' \sin\delta]^2}{\Delta}
 \end{array} \right) \, ,
\end{align}
with $\delta = \pi\eta$. Neither $E_s^-$ nor $M_s$ depend on $s'$, since the pockets are symmetric with respect to the $k_x$ axis.

Since the matrices $M_\pm$ are symmetric, the quadratic form \eqref{eq: disp low doping} can be diagonalized by a rotation in momentum space, yielding
\begin{equation} \label{eq: disp low doping 2}
 E_\kk^- = E_s^-
 - \frac{\tilde k_\perp^2}{2m_{s\perp}} - \frac{\tilde k_\parallel^2}{2m_{s \parallel}} \, ,
\end{equation}
where $m_{s\parallel}$ and $m_{s\perp}$ are the two eigenvalues of $M_s$, while $\tilde k_\parallel$ and $\tilde k_\perp$ are the momentum coordinates in the rotated basis.
We will show below that the incommensurability $\eta$ is proportional to the doping $p$ at low hole doping. Hence, to leading order in our expansion, we can sometimes replace the mass matrix by its value for $\eta=0$, where it is the same for all pockets.

Within the quadratic expansion \eqref{eq: disp low doping}, the hole pockets in the lower quasiparticle band $E_\kk^-$ have the form of an ellipse.
The number of holes accommodated in each pocket is given by the area of the ellipse divided by $(2\pi)^2$, that is,
\begin{equation}
 p_s = \frac{1}{2\pi} \sqrt{m_{s\perp} m_{s\parallel}} \, |E_s^-| \, .
\end{equation}
%


\subsubsection{Incommensurability}

Assuming that the N\'eel state at half-filling is replaced by a spiral state upon arbitrarily small hole-doping, we now show that the incommensurability $\eta$ is proportional to the doping $p$ for small $p$, and that only two (not four) hole pockets are present.
To this end we consider the minimization of the ground state energy
\begin{equation}
 E(\QQ) = \int_\kk [E_\kk^-(\QQ) + \mu(\QQ)] f[E_\kk^-(\QQ)] + \frac{\Delta^2}{U}
\end{equation}
for fixed density with respect to the ordering wave vector $\QQ$.
Here we have made the $\QQ$ dependence of all involved quantities explicit.
The second contribution to the integral, involving the chemical potential, yields simply $\mu(\QQ) n$.
The energy can also be written in the form
\begin{equation}
 E(\QQ)  = \int_\kk \tilde E_\kk^-(\QQ) -
 \sum_{ss'} \int_{\kk \in P_{ss'}} \tilde E_\kk^-(\QQ) +\frac{\Delta^2}{U} \, ,
\end{equation}
where $\tilde E_\kk^-(\QQ) = E_\kk^-(\QQ) + \mu(\QQ)$, and $P_{ss'}$ denotes the set of momenta contained in the pocket labelled by $s$ and $s'$.
Expanding the energy as a function of $\QQ$ around $\boldsymbol{\pi} = (\pi,\pi)$ to second order in $\eta$ yields
\begin{eqnarray} \label{eq: energy1}
 E(\QQ) &=&
 \int_\kk \tilde E_{\kk}^-(\boldsymbol{\pi}) +\frac{\Delta^2}{U}+
 \int_\kk \frac{1}{2} \left. \frac{\partial^2 \tilde E_\kk^-(\QQ)}{\partial\eta^2}
 \right|_{\eta=0} \eta^2 \nonumber \\
 &+& 2\pi \sum_s \frac{p_s^2}{\sqrt{m_{s\perp} m_{s\parallel}}} \nonumber \\
 &+& 2\Delta (p_+ + p_-) - 4\pi t (p_+ - p_-) \eta \, .
\end{eqnarray}
Here in the expansion, we considered that the partial derivative with respect to $\eta$ of the energy equals its total derivative, since the gap and the chemical potential, implicitly depending on $\eta$, minimize the ground state energy. 

The prefactor of the term proportional to $\eta^2$ in Eq.~\eqref{eq: energy1} is essentially the stiffness at half-filling,
\begin{equation}
 \frac{1}{2} \int_\kk \left. \frac{\partial^2 \tilde E_\kk^-(\QQ)}{\partial\eta^2}
 \right|_{\eta=0} =
 - \frac{\pi^2}{2} \int_\kk \gamma^x_{\kk}\gamma^x_{\kk+\QQ} \frac{\Delta^2}{e_{\kk}^3}
 = 2\pi^2 J(n=1) \, .
\end{equation}
At half-filling, $\QQ=(\pi,\pi)$ and $J^{\perp}_{\alpha\beta}=J^{\Box}_{\alpha\beta}=J \delta_{\alpha\beta}$.
In the third term in Eq.~\eqref{eq: energy1} one can replace the masses by their values at half-filling, so that the pocket index $s$ can be dropped. The $\eta$ dependence of the masses yields a contribution beyond quadratic order (in the doping) to the energy.
The energy~\eqref{eq: energy1} can then be written as
\begin{eqnarray}
 E(\QQ) &=& E(\boldsymbol{\pi}) + 2\pi^2 J(n=1) \, \eta^2 +
 \frac{2\pi}{\sqrt{m_\parallel m_\perp}} (p_+^2 + p_-^2) \nonumber \\
 &+& 2\Delta (p_+ + p_-) - 4\pi t (p_+ - p_-) \eta \, .
\end{eqnarray}
Minimizing the energy with respect to $\eta$ yields
\begin{equation} \label{eq: eta vs p}
 \eta = \frac{t}{\pi J(n=1)} (p_+ - p_-) \, ,
\end{equation}
and
\begin{eqnarray}
 E(\QQ) &=& E(\boldsymbol{\pi}) +
 \left( \frac{\pi}{ \sqrt{m_\perp m_\parallel}}
 - \frac{2t^2}{J(n=1)} \right) \left( p_+ - p_- \right)^2 \nonumber \\
 &+& \frac{\pi}{\sqrt{m_\perp m_\parallel}} \left( p_+ + p_- \right)^2
\end{eqnarray}
Since the prefactor of the term of order $(p_+ - p_-)^2$ is negative, the energy at fixed total hole density $p = 2(p_+ + p_-)$ is minimized when the volume difference $p_+ - p_-$ is maximal, that is, when only the two pockets centered around
$\left(\frac{\pi}{2} - \pi\eta,\pm \frac{\pi}{2} \right)$ are present, so that $p_- = 0$. The relation \eqref{eq: eta vs p} then yields the simple linear relation
\begin{equation} \label{eq: eta doping}
 \eta = \frac{t}{2\pi J(n=1)} \, p
\end{equation}
between the doping $p$ and the incommensurability $\eta$.
In the low hole doping limit this result agrees perfectly with the numerical results for the doping dependence of $\eta$ shown in Fig.~\ref{fig: eta}. 

An approximate version of Eq.~\eqref{eq: eta doping} was also found in the low doping limit in Ref.~\cite{Chubukov1995} within a strong coupling approximation. Furthermore, a similar formula was also derived in the context of the $t$-$J$ model in the Refs.~\cite{Shraiman1992,Sushkov2009}.

From neutron-scattering measurements of spin fluctuations in La$_{2-x}$Sr$_x$CuO$_4$, Yamada {\em et al.}~\cite{Yamada1998} extracted the approximate empiric relation $\eta=p$ between the incommensurability and the hole-doping. For our choice of parameters we have $J(n=1) \approx 0.18t$, leading to $\eta \approx 0.88 p$, which happens to be quite close to the Yamada relation. However, the prefactor of this linear relation is obviously not universal. A recent slave boson mean field study of the Hubbard model yielded numerical results consistent with the empirical Yamada relation in the intermediate to strong coupling regime \cite{Klett2024}.


\subsubsection{In-plane stiffness at low doping}

In this section we analyse the difference between the spin stiffness at half-filling $J(n=1)$ and the low hole doping limit of the in-plane stiffness
$J_{\alpha\alpha}^\Box(n \to 1^-)$.
All contributions in Eq.~\eqref{eq: Jin} are regular in the low doping limit, apart from the contribution containing $f'(E^-_\kk)$ to the first (bare) term, see Eq.~\eqref{eq: Jin0}. This term leads to the discontinuity
\begin{equation} \label{eq: Jin diff1}
 J(n=1) - J_{\alpha\alpha}^\Box(n \to 1^-) =
 \Delta^2 \int_\kk \gamma_\kk^\alpha \gamma_{\kk+\QQ}^\alpha
 \frac{f'(E^-_{\kk})}{4e_\kk^2} \, .
\end{equation}
Since $f'(E) = -\delta(E)$ at $T=0$, the only contributions to this integral come from momenta on the Fermi surface of the (two) hole pockets centered around
$\kk_{+s'} = (\frac{\pi}{2} - \pi\eta,s'\frac{\pi}{2})$.
For small hole doping, the pockets are small, and $\eta \to 0$, so that we can replace
$\gamma_\kk^\alpha \gamma_{\kk+\QQ}^\alpha/e_\kk^2 \to
 \gamma_{\kk_{+s'}}^\alpha \gamma_{\kk_{+s'}+\QQ}^\alpha/e_{\kk_{+s'}}^2 \to - 4t^2/\Delta^2$.
In the last step a dispersion of the form Eq.~\eqref{eq: dispersion} was assumed.
With the elementary integral over the two pocket Fermi surfaces,
$\int_\kk f'(E^-_\kk) = - \frac{1}{\pi} \sqrt{m_\parallel m_\perp}$,
we then obtain
\begin{equation} \label{eq: jump1}
 J(n=1) - J_{\alpha\alpha}^\Box(n \to 1^-) =
 \frac{t^2}{\pi} \sqrt{m_\perp m_\parallel} \, ,
\end{equation}
where $m_\perp$ and $m_\Box$ are the effective masses characterizing the dispersion of the two hole pockets in the limit $n\rightarrow 1^-$.
For a dispersion of the form Eq.~\eqref{eq: dispersion}, with $t'<0$, one obtains \cite{Metzner2019}
\begin{equation} \label{eq: masses}
 m_\parallel = - \frac{1}{4t'} \, , \quad
 m_\perp = \frac{\Delta}{8t^2 + 4\Delta t'} \, .
\end{equation}
For the parameters used for the stiffnesses shown in Fig.~\ref{fig: J_groundstate}, Eq.~\eqref{eq: jump1} yields $J(n=1) - J_{\alpha\alpha}^\Box(n \to 1^-) \approx 0.094t$, which agrees with the numerical results.
Note that $m_\parallel$ in Eq.~\eqref{eq: masses} diverges for $t' \to 0$. This indicates, once again, that the case of pure nearest neighbor hopping is very special.


\subsubsection{Out-of-plane stiffness at low doping}

We now apply a similar analysis to the out-of-plane stiffness, as given by Eq.~\eqref{eq: Jout}.
At first sight one might expect that, as before, the discontinuity at half-filling is determined by the term containing the factor $f'(E^-_\kk)$.
However, this term actually vanishes in the low doping limit $n \to 1^-$, since the regular prefactor of $f'(E^-_\kk)$ in Eq.~\eqref{eq: Jout} vanishes for $\kk \to \kk_{+s'} \to (\frac{\pi}{2},s'\frac{\pi}{2})$ and $\QQ \to (\pi,\pi)$.
The only singular contribution is the one in the first line of Eq.~\eqref{eq: Jout} for $l=l'=-$,
\begin{eqnarray} \label{eq: Jout1}
 J_{\alpha\alpha}^{\perp,1} &=& \frac{1}{4} \int_{\bs{k}}
 \left(1+\frac{h_\kk}{e_\bs{k}} \right)
 \left(1-\frac{h_{\kk+\QQ}}{e_{\kk+\QQ}}\right)  \nonumber \\
 &\times& \left( \gamma^\alpha_{\kk+\QQ} \right)^2
 \frac{f(E^-_\kk) - f(E^{-}_{\kk+\QQ})}{E^-_\kk - E^{-}_{\kk+\QQ}} \, .
\end{eqnarray}
For the N\'eel \ state, $E^-_{\bs{k}} = E^-_{\bs{k}+\bs{Q}}$ for all $\kk$,
so that $J_{\alpha\alpha}^{\perp,1}$ becomes
\begin{eqnarray}
 \left. J_{\alpha\alpha}^{\perp,1} \right|_{\rm \QQ=(\pi,\pi)} &=&
 \frac{1}{4} \int_\kk \left( \gamma^\alpha_{\kk+\QQ} \right)^2
 \left( 1 + \frac{h_\bs{k}}{e_{\bs{k}}} \right)
 \left( 1 - \frac{h_{\bs{k}+\bs{Q}}}{e_{\bs{k}+\bs{Q}}} \right)  \nonumber \\
 &\times& f'(E^-_{\bs{k}}) \, .
\end{eqnarray}
At half-filling this term vanishes, since $E_\kk^- < 0$ for all $\kk$ (no Fermi surface).
Hence, the discontinuity is given by
\begin{equation} \label{eq: perp discont}
 J(n=1) - J_{\alpha\alpha}^\perp(n \to 1^-) = - J_{\alpha\alpha}^{\perp,1}(n \to 1^-) \, .
\end{equation}

Due to the difference of Fermi functions, the integrand in Eq.~\eqref{eq: Jout1} is non-zero only if $\kk$ or $\kk+\QQ$ are situated in one of the hole pockets. Let us first evaluate the former contribution. In the limit $n \to 1^-$, the size of the pockets shrinks to zero, and the regular functions in the integrand of Eq.~\eqref{eq: Jout1} can replaced by their values at the pocket center $\kk_{+s'}$, that is, $h_\kk \to 0$, $h_{\kk+\QQ} \to 0$, $e_\kk \to \Delta$, $e_{\kk+\QQ} \to \Delta$, $\gamma_{\kk+\QQ}^\alpha \to -2t$, and $E_\kk^- - E_{\kk+\QQ}^- \to 4\pi\eta t$.
The integral over the difference of Fermi functions yields simply the volume of the hole pockets, that is, the doping $p$. Collecting all the factors in the limit $\kk \to \kk_{+s'}$ as listed above, one obtains the first contribution
$J_{\alpha\alpha}^{\perp,1a} = - \frac{t}{4\pi\eta} p$ to the integral in Eq.~\eqref{eq: Jout1}.
The second contribution, $J_{\alpha\alpha}^{\perp,1b}$, from momenta $\kk+\QQ$ inside a hole pocket, is obtained analogously and has the same value. We thus find
\begin{equation}
 J_{\alpha\alpha}^{\perp,1} = - \frac{t}{2\pi\eta} p = - J(n=1) \, ,
\end{equation}
where the last relation follows from Eq.~\eqref{eq: eta doping}. Hence, the discontinuity of the out-of-plane stiffness at half-filling is given by the stiffness at half-filling itself, so that
\begin{equation} \label{eq: J out zero}
 J_{\alpha\alpha}^\perp(n \to 1^-) = 0 \, .
\end{equation}
We emphasize that this striking result hinges on the relation \eqref{eq: eta doping} between the doping and the incommensurability.

\subsection{$t'$ dependence}
\label{sec: tprime dependence}

It is instructive to study the dependence on the next-nearest-neighbor hopping amplitude $t'$, which was set to $t'=-0.2t$ in the preceding sections. We now vary $t'$ to assess its effect on the magnetic stiffnesses and the validity of our low-doping expansion. We fix the coupling value as before to $U=2.35$. We first observe that, as $t'$ approaches zero, the hole pockets become narrower. In fact, the expansion~\eqref{eq: disp low doping 2} fails in the $t'\rightarrow 0$ limit due to the divergence of the parallel mass parameter $m_{\parallel}$ (see Eq.~\eqref{eq: masses}). Nevertheless, for any small but finite value of $t'$, there always exists a doping regime, as $n\rightarrow 1^-$, where the expansion~\eqref{eq: disp low doping 2} remains valid, and the hole pockets are accurately represented by small (albeit highly eccentric) ellipses. 

\begin{figure}[t]
    \centering
    \includegraphics[width=0.48\textwidth]{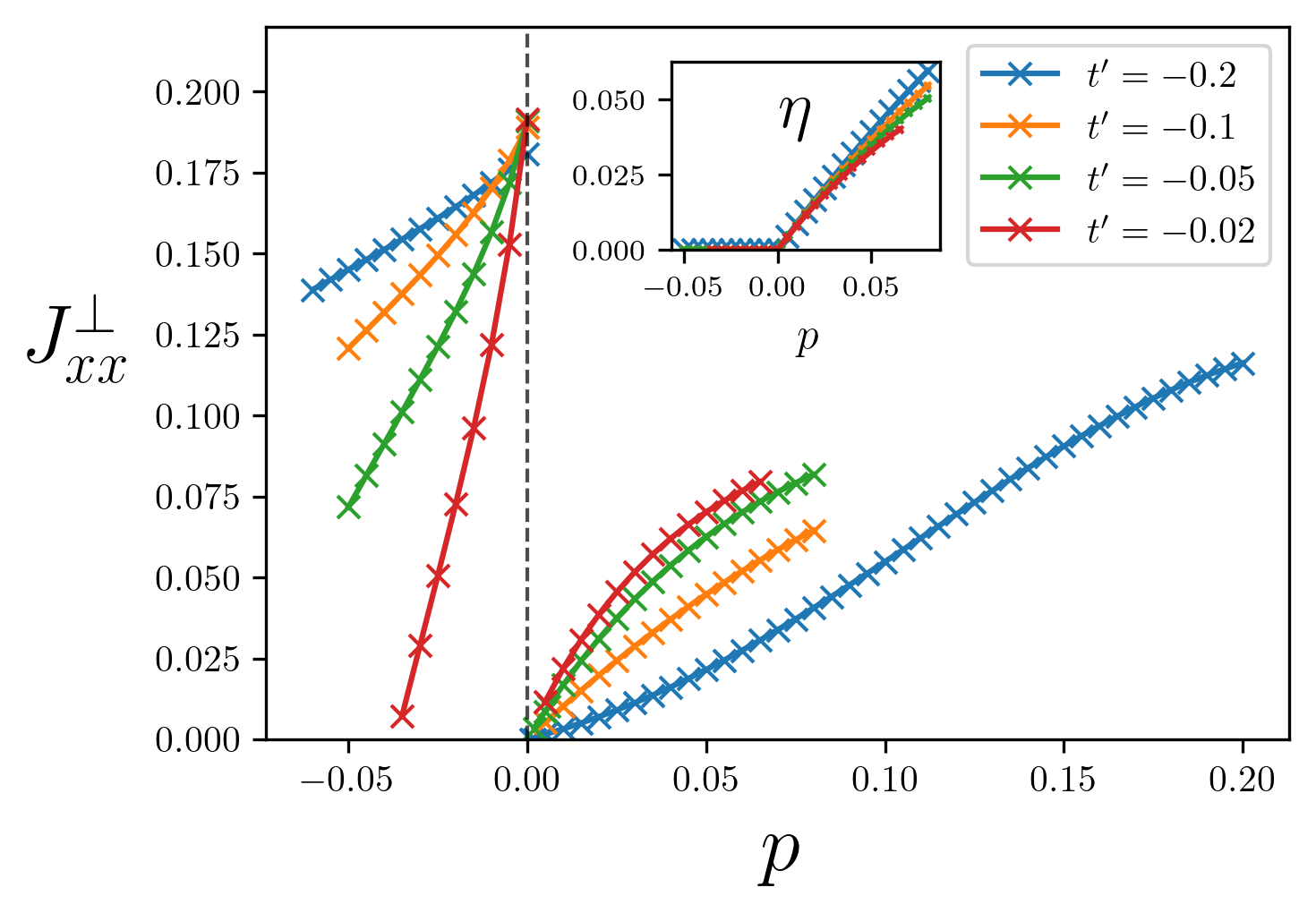}
    \caption{Doping dependence of out-of-plane stiffnesses for various $t'$ values. In the inset, the incommensurability $\eta$ as a function of the doping for various $t'$. }
    \label{fig: J out vs tprime}
\end{figure}

In Fig.~\ref{fig: J out vs tprime}, we show the out-of-plane stiffness $J^\perp_{xx}$ as a function of doping $p$ for various values of $t'$. As a general feature, we observe stable N\'eel order at half-filling and upon electron-doping, which transitions to spiral order for any finite small hole-doping. The resulting incommensurability $\eta$ is displayed in the inset. The linear dependence of $\eta$ in the low doping limit is almost independent of the value of $t'$. Specifically, its dependence on doping is consistently given by Eq.~\eqref{eq: eta vs p}, and the stiffness at half-filling only has a very weak dependence on $t'$. The doping behavior of the stiffness on the electron-doped side is strongly $t'$-dependent. On the hole-doped side, all curves approach zero independently of $t'$, which confirms the generality of our finding in Eq.~\eqref{eq: J out zero}.

\begin{figure}[t]
    \centering
    \includegraphics[width=0.48\textwidth]{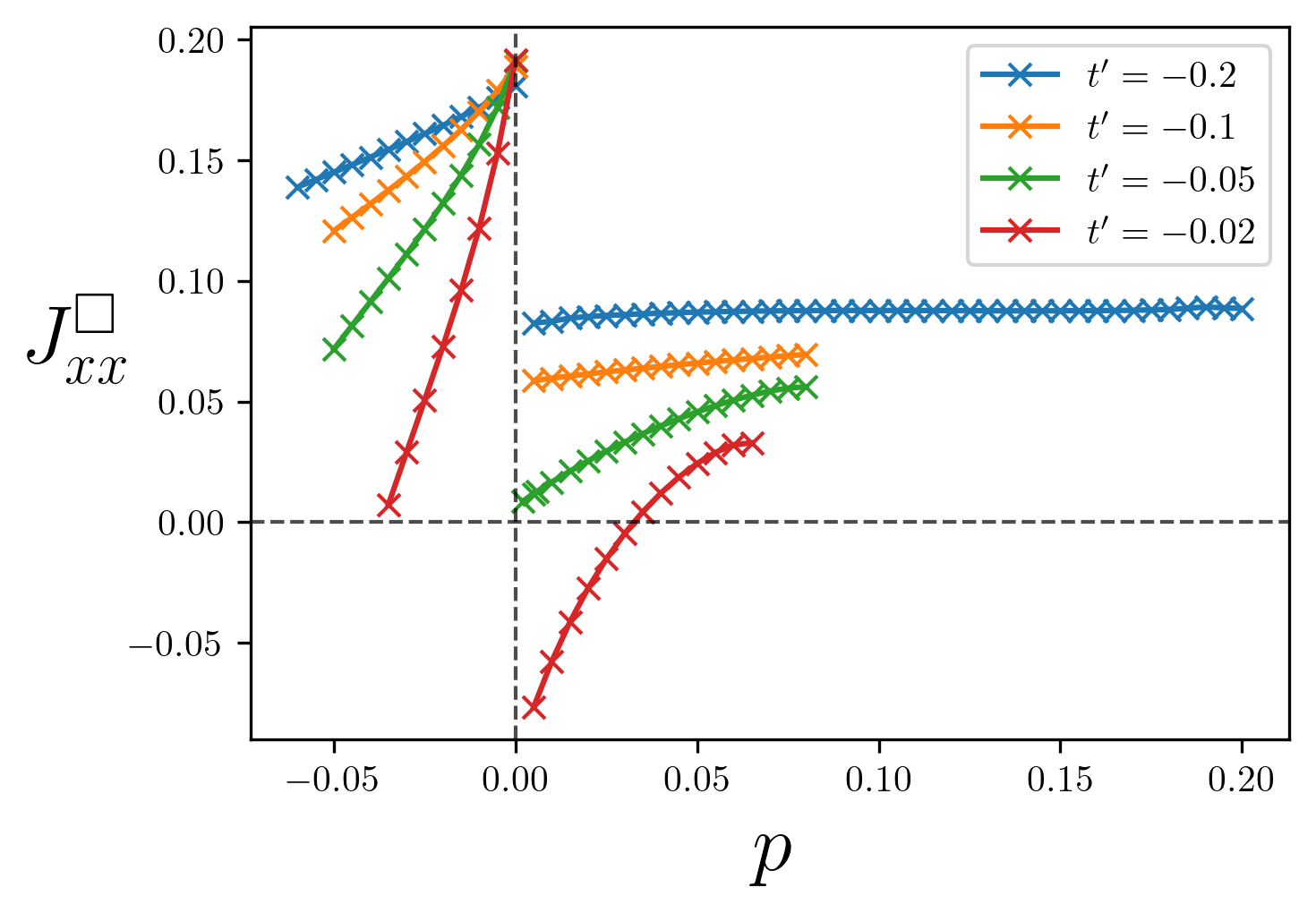}
    \caption{Doping dependence of in-plane stiffnesses for various $t'$ values. }
    \label{fig: J in vs tprime}
\end{figure}

In Fig.~\ref{fig: J in vs tprime}, we plot the in-plane stiffness $J^{\smsqr}_{xx}$. On the hole-doped side, we observe, as before, a sharp jump whose magnitude sizably depends on the $t'$ value. We confirmed that the analytical prediction in Eq.~\eqref{eq: Jin diff1} remains valid across all cases. This strong $t'$ dependence of $J^{\smsqr}_{xx}$ in the low-doping limit for small $|t'|$ is analytically justified by considering the $t'\rightarrow 0$ limit: the parallel mass parameter $m_\parallel$ (see Eq.~\eqref{eq: masses}) diverges, which, as a direct consequence, causes the jump in Eq.~\eqref{eq: Jin diff1} to also diverge. Indeed, for a small value like $t'=-0.02t$, the jump is already large enough to drive the in-plane stiffness to negative values. This negative stiffness implies that the spiral order  with $\QQ=(\pi,\pi-2\pi\eta)$ is not stable in this regime, and another mean-field state, specifically a diagonal spiral order with $\QQ = (\pi - 2\pi\eta,\pi-2\pi\eta)$, is found to have a lower energy. Alternatively, since we observe a diverging charge susceptibility, it is also possible that a spin-charge stripe state has the lowest energy in this regime \cite{Scholle2023}. 

As a general remark, this analysis illustrates that in the low doping limit there will always be a finite small $|t'|$ value below which the planar spiral order becomes unstable, necessitating a search for other states, such as diagonal spiral or stripe order. Conversely, for $|t'|$ values larger than the range explored here, specifically for $-0.23t < t' < -0.27t$, the N\'eel state remains stable only for very small doping, and the system becomes entirely paramagnetic for even larger $|t'|$. We emphasize that the specific boundaries of these magnetic phases depend on the coupling $U$, as explained above. 

\subsection{Temperature dependence}
\label{sec: temperature dependence}

\begin{figure}[t]
    \centering
    \includegraphics[width=0.48\textwidth]{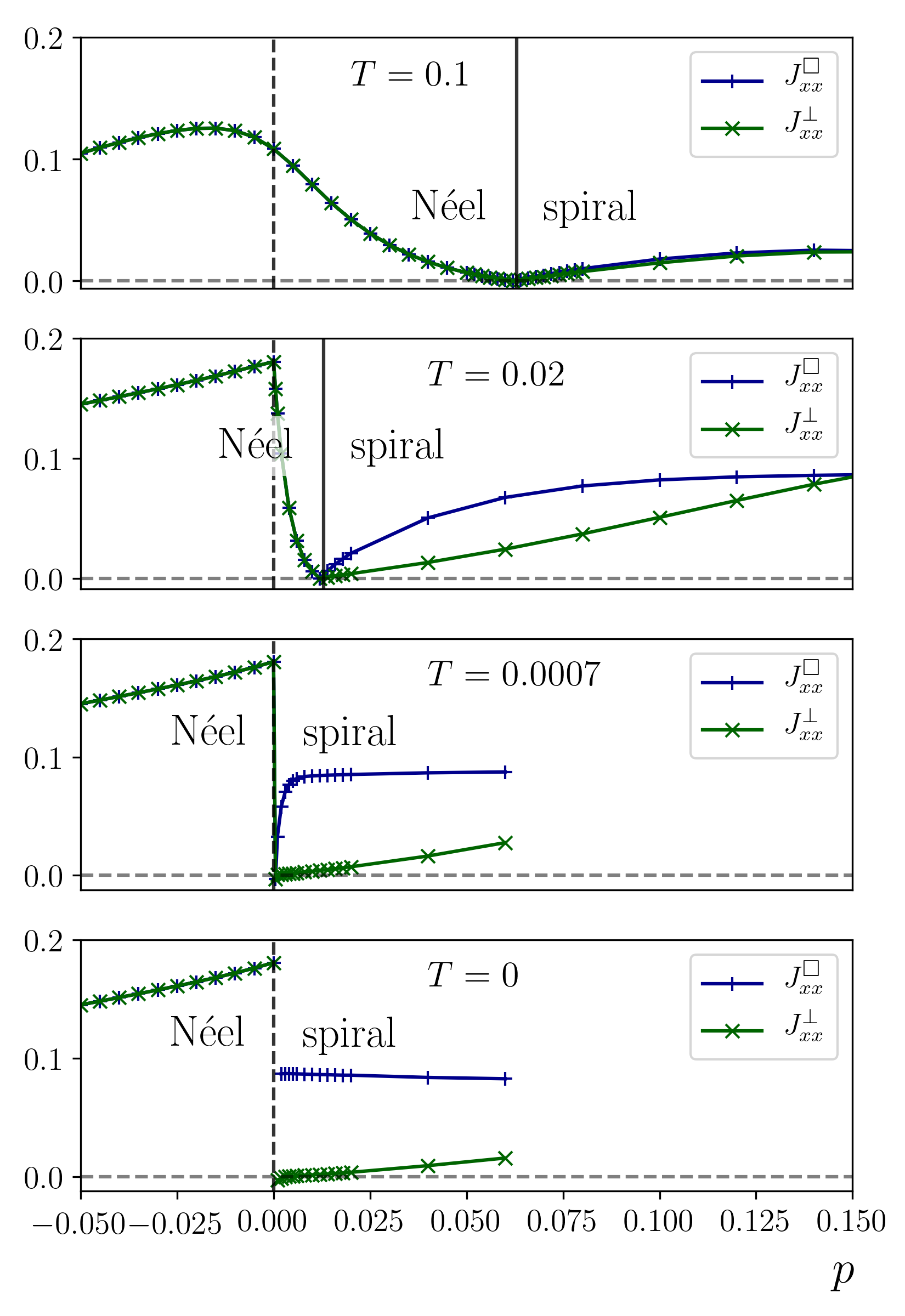}
    \caption{Doping dependence of in-plane and out-of-plane stiffnesses for various temperatures.}
    \label{fig: J vs T}
\end{figure}

So far, we focused on the ground state, where many momentum integrals can be analytically evaluated in the low doping limit. It is instructive to analyze the behavior of the stiffnesses at the N\'eel-spiral transition also at finite temperature.
At finite $T$ the N\'eel-spiral transition occurs at a finite critical doping $p_\mathrm{c}$, whose value is temperature dependent and approaches zero only for $T \to 0$. The incommensurability $\eta$ is always a smooth function of doping, at any temperature.

In Fig.~\ref{fig: J vs T} the spatial in-plane and out-of-plane stiffnesses are shown as functions of doping for various temperatures. In the spiral regime only the $xx$ component is shown. The $yy$ component behaves similarly, and the mixed $xy$ component vanishes.
Both the out-of-plane and in-plane stiffnesses vanish at the transition, at any finite temperature. The same behavior was already observed for the stiffnesses as obtained from a dynamical mean-field calculation at a sizable finite temperature, and it was explained by the merging of the two Goldstone poles at $\pm\QQ$ at the N\'eel ordering vector $(\pi,\pi)$ as one approaches the transition from the spiral regime \cite{Goremykin2024}.

In view of the distinct behavior we found for the ground state, it is useful to work out the argument of the merging poles in some detail. Let us start with the out-of-plane stiffness, which is obtained from the 33-component of the static susceptibility $\chi^{33}(\qb) = \chi^{33}(\qb,0)$. Close to the N\'eel-spiral transition, the inverse of $\chi^{33}(\qb)$ can be expanded around the N\'eel wave vector $(\pi,\pi)$ as
\begin{equation} \label{eq: chi33 exp}
 [\chi^{33}(\pi+\delta q_x,\pi)]^{-1} =
 \alpha(n) \delta q_x^2 + \beta(n) \delta q_x^4 \, ,
\end{equation}
where $\alpha(n)$ and $\beta(n)$ are density dependent coefficients.
Its minimum is located at $\delta{\bar q}_x = 0$ for $\alpha(n) > 0$, corresponding to a N\'eel state, and at $\delta{\bar q}_x = [- \frac{1}{2} \alpha(n)/\beta(n)]^{1/2}$ for $\alpha(n) < 0$, corresponding to a spiral state. The stiffness is given by $J = m^2 \alpha(n)$ in the N\'eel regime, and by
\begin{equation}
 J_{xx}^\perp =
 \frac{1}{2} m^2 \left. \frac{d^2 (\chi^{33})^{-1}}{d \delta q_x^2}
 \right|_{\delta{\bar q}_x} = - 2m^2 \alpha(n)
\end{equation}
in the spiral regime. Assuming that $\alpha(n)$ is a continuous function of density, we see that the stiffnesses vanish continuously along with $\alpha(n)$ on both sides of the N\'eel-spiral transition.
This explanation holds only when all relevant quantities, such as the gap, the incommensurability, and the Fermi surface, are continuous functions of the doping, which is no longer true at zero temperature.

The above arguments hold analogously for the in-plane stiffness. The Goldstone poles of the corresponding susceptibility $\chi^{22}(\qb)$ are also located at $\pm\QQ$, which are distinct points in the spiral state, while they merge at the N\'eel-spiral transition.

At $T=0$, the above continuity argument breaks down since the coefficients in Eq.~\eqref{eq: chi33 exp} can exhibit jumps, and the stiffnesses exhibit discontinuities at the N\'eel-spiral transition. That the out-of-plane stiffness nevertheless vanishes on the spiral side of the transition has distinct (and more subtle) reasons, which have been explained in Sec.~\ref{sec: low doping}.


\subsection{\Neel-spiral transition at finite doping}

\begin{figure*}[t]
    \centering
    \includegraphics[width=0.95\textwidth]{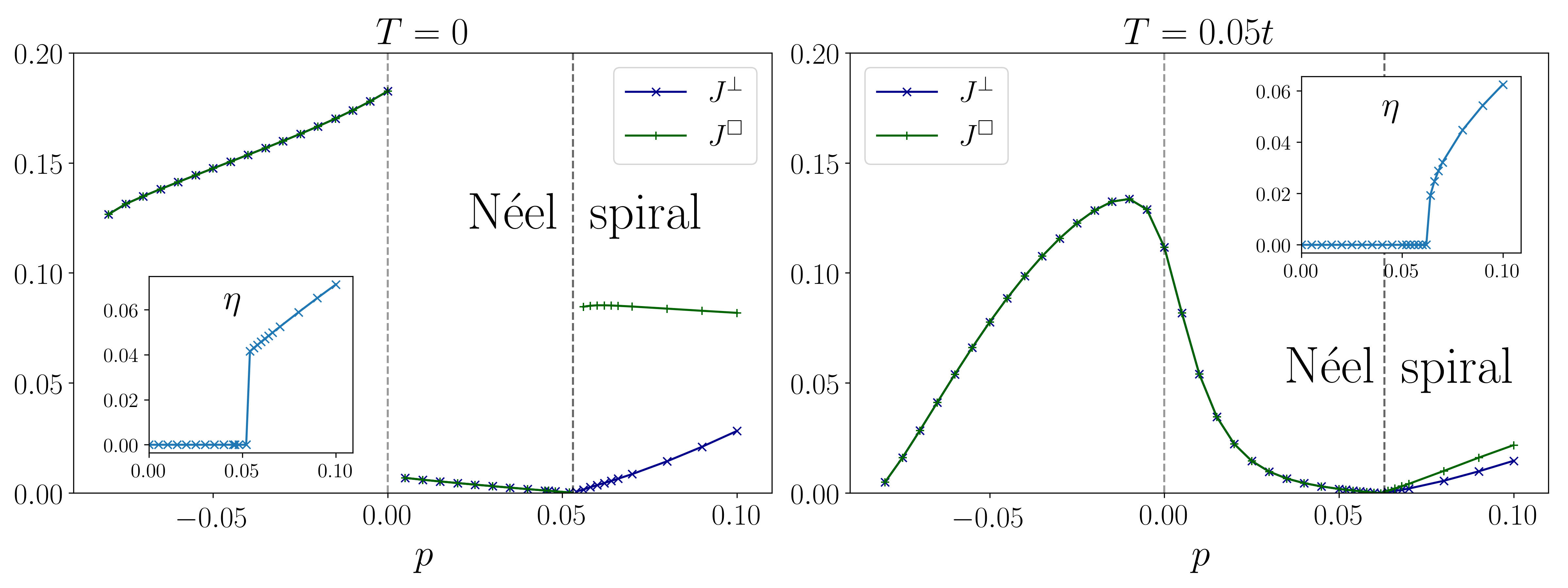}
    \caption{In-plane $J^{\smsqr}$ and out-of-plane $J^{\perp}$ stiffnesses as functions of doping for $U=2.2t$. Left panel: $T=0$, right panel: $T=0.05t$. The insets show the doping dependence of the incommensurability $\eta$.}
    \label{fig: Neel spiral finite doping}
\end{figure*}

In the previous sections, we focused on the case where in the ground state the N\'eel-spiral transition occurs at half-filling. This situation arises, at least in mean-field theory, when the Hubbard interaction is sufficiently strong \cite{Chubukov1992}.
For weaker interactions, the N\'eel state may remain stable for low hole doping, and the N\'eel-spiral transition occurs at a finite critical hole doping $p_c$.
In our mean-field theory, we can realize this case by slightly reducing the coupling value to $U=2.2t$, keeping the value of $t'$ as before. For these parameters, the transition occurs at $p_c \approx 0.053$.

In Fig.~\ref{fig: Neel spiral finite doping} we show the corresponding in-plane and out-of-plane stiffnesses in the ground state and at a finite temperature. At $T=0.05t$, shown on the right panel, the incommensurability $\eta$ is continuous at the transition, such that the transition is second order and both stiffnesses vanish at the transition, as explained in the previous section by the merging of the Goldstone points. The situation is however different at zero temperature, shown on the left panel, where the transition is first order, with a jump in the incommensurability $\eta$. Here, there is no \emph{a priory} reason for the stiffnesses to vanish, and indeed the in-plane stiffness jumps at the critical doping. However, the out-of-plane stiffness still approaches zero at the transition, at least within the numerical accuracy, which is surprising.

\begin{figure}[t]
\centering
 \includegraphics[width=0.48\textwidth]{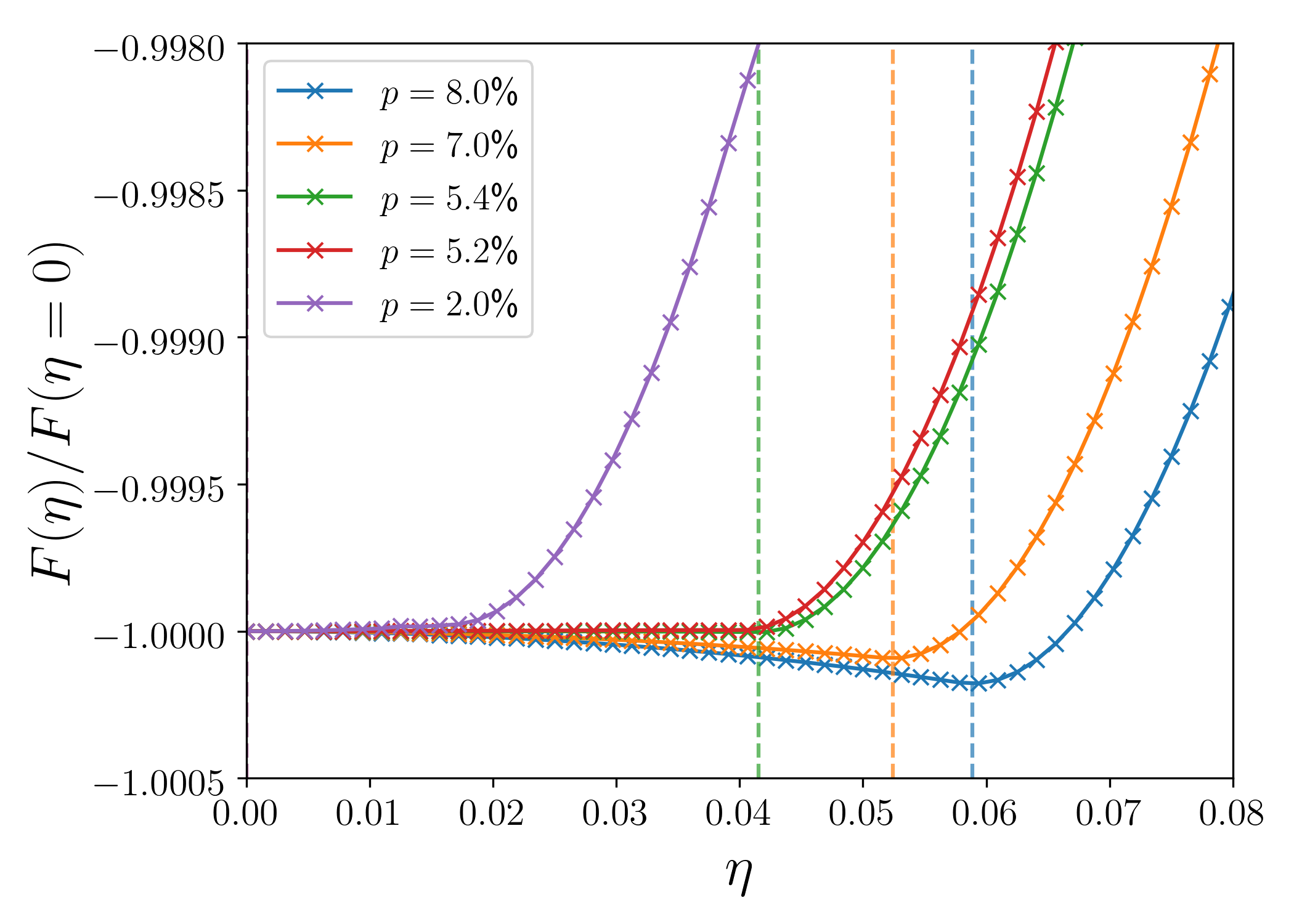}
 \caption{Landau energy as a function of the incommensurability $\eta$ at $T=0$ for various choices of the doping $p$.}
 \label{fig: Free Energy}
\end{figure}
To gain further insight, we show the Landau energy as a function of the incommensurability $\eta$ for various choices of the hole doping in Fig.~\ref{fig: Free Energy}. One can see that the $Q_x$-dependence becomes very flat for doping values near the critical doping $p_c$. In particular, the curvature at $Q_x = \pi$ is practically zero, as if the transition was of second order. In view of this situation, the vanishing of the N\'eel stiffness and the out-of-plane spiral stiffness is plausible.
The distinct behavior of the in-plane stiffness can again be traced back to the term involving $f'(E_\kk^-)$ in Eq.~\eqref{eq: Jin0}, which yields a contribution proportional to the number of hole pockets (see the discussion leading to Eq.~\eqref{eq: jump1}). This number changes abruptly from four to two at the discontinuous N\'eel-spiral transition.
When going beyond mean-field theory, the flatness of the Landau energy for doping values near $p_c$ can be expected to yield particularly strong fluctuation effects.

In the low doping limit, where the N\'eel state is now stable, we can analytically compute the size of the drop of the stiffness at half-filling and $T=0$ as before. In the N\'eel state the in-plane and out-of-plane stiffnesses coincide, that is, there is only one stiffness $J$. It is most convenient to use the formulae for the in-plane stiffness, \eqref{eq: Jin} and \eqref{eq: Jin0}, with $\QQ = (\pi,\pi)$, to compute $J$. The only discontinuous contribution contains the term $f'(E^-_\kk)$, leading to the discontinuity \eqref{eq: Jin diff1}. While the spiral state has only two hole pockets at low doping, the N\'eel state has four of them, yielding an extra factor two compared to Eq.~\eqref{eq: jump1}, such that
\begin{equation}
 J(n=1) - J(n\rightarrow 1^-) = \frac{2t^2}{\pi}\sqrt{m_{\perp}m_{\parallel}}
\end{equation}
Inserting the effective masses $m_\perp$ and $m_\parallel$ from Eq.~\eqref{eq: masses}, we find $J(n=1) - J(n\rightarrow 1^-) \approx 0.175t$, in agreement with the numerical results.


\section{Stability of magnetic order} \label{sec: III}

We now investigate whether quantum fluctuations destroy the magnetic order of the ground state, obtained within mean-field theory, in the lightly hole-doped regime.
We first review the analysis of this issue via a quantum non-linear sigma model, from which emerges the need for a more comprehensive approach.


\subsection{Non-linear sigma model}

Low-energy small-momentum order parameter fluctuations of a spiral magnet can be described by a quantum non-linear sigma model of the form \cite{Bonetti2022gauge}
\begin{equation} \label{eq: spiral nlsm}
 {\cal S}[{\cal R}] = \frac{1}{2} \int dx \, {\rm Tr} \left[
 {\cal P}_{\mu\nu} (\partial_\mu {\cal R}^T) (\partial_\nu {\cal R}) \right] \, ,
\end{equation}
where ${\cal R}(x)$ with $x=({\bf r},\tau)$ is a SO(3) matrix depending on the position $\bf r$ (in continuous space) and on the imaginary time $\tau$, while $\partial_\mu$ with $\mu \in \{\tau,x,y\}$ denotes partial derivatives with respect to time or space.
The coefficient in this action is a $3\times3$ matrix of the form
${\cal P}_{\mu\nu} = \frac{1}{2} {\rm Tr}[{\cal J}_{\mu\nu}] 1_3 - {\cal J}_{\mu\nu}$,
where $1_3$ is the three-dimensional unit matrix, and ${\cal J}_{\mu\nu}$ is a diagonal $3\times3$ matrix \cite{factor2},
\begin{equation}
 {\cal J}_{\mu\nu} =
 {\rm diag}\left( \textstyle{\frac{1}{2}} J_{\mu\nu}^\perp,
 \textstyle{\frac{1}{2}} J_{\mu\nu}^\perp, J_{\mu\nu}^\Box \right) \, .
\end{equation}
Here $J_{\tau\tau}^a$ with $a \in \{\Box,\perp\}$ are the in-plane and out-of-plane temporal stiffnesses $Z^a$, while $J_{\alpha\beta}^a$ with $\alpha,\beta \in \{x,y\}$ are the corresponding spatial stiffnesses.

In the case of N\'eel order, the corresponding non-linear sigma model assumes the simpler form \cite{Haldane1983,Haldane1983a}
\begin{equation} \label{eq: Neel nlsm}
 {\cal S}[\hat n] = \frac{1}{2} \int dx \, \left(
 Z|\partial_\tau \hat n|^2 + J |\nabla \hat n|^2 \right) \,
\end{equation}
where $\hat n(x)$ is a unit vector fluctuating in space and time.

The non-linear sigma model \eqref{eq: spiral nlsm} has a convenient CP$^1$ representation, where the matrix ${\cal R}(x)$ is expressed in terms of a spinor of two complex Schwinger bosons
$z(x) = (z_\up(x),z_\down(x))$ subject to the local constraint
$|z_\up(x)|^2 + |z_\down(x)|^2 = 1$.
The CP$^1$ action reads \cite{Chubukov1994, Bonetti2022gauge}
\begin{equation} \label{eq: cp1 action}
 {\cal S}[z,z^*] = \int dx \left[
 J_{\mu\nu}^\perp (\partial_\mu z^*)(\partial_\nu z) -
 (J_{\mu\nu}^\perp - 2J_{\mu\nu}^\Box) j_\mu j_\nu \right] \, ,
\end{equation}
with $j_\mu = \frac{1}{2} \left[ z^* (\partial_\mu z) - (\partial_\mu z^*) z \right]$.
For the N\'eel case one obtains the same expression with $J_{\mu\nu}^\perp \to 2J_{\mu\nu}$ and $J_{\mu\nu}^\Box = 0$.

The non-linear sigma model is spin rotation invariant. At finite temperatures, the SU(2) symmetry cannot be broken in two space dimensions. At zero temperature, however, a spontaneous breaking of the SU(2) symmetry is possible. In the CP$^1$ representation, SU(2) symmetry breaking is reflected by a condensation of the Schwinger bosons, that is, by a non-zero expectation value $\langle z_\sigma \rangle$.

In a saddle point approximation, which becomes exact in a large $N$ extension of the CP$^1$ action, the condensate fraction $n_0$ is determined by the equation \cite{Chubukov1994, Auerbach1994, Bonetti2022gauge}
\begin{equation} \label{eq: spingap equation}
 n_0 + T \sum_{\omega_n} \int_\qb
 \frac{2}{Z^\perp \omega_n^2 + J_{\alpha\beta}^\perp q_\alpha q_\beta + Z^\perp m_s^2}
 = 1 \, ,
\end{equation}
where $\omega_n = 2\pi nT$ are bosonic Matsubara frequencies, and $m_s$ is a dynamically generated spin gap. Note that only the out-of-plane modes contribute to the saddle point solution. By the Goldstone theorem, $m_s = 0$ if $n_0$ is non-zero.
At $T > 0$, the condition \eqref{eq: spingap equation} cannot be fulfilled for $m_s = 0$, since the classical contribution from $\omega_n = 0$ leads to an infrared divergent momentum integral. Hence, $n_0 = 0$ and $m_s > 0$ in this case, in agreement with the Mermin-Wagner theorem.

In the ground state, the Matsubara sum is an integral over a continuous frequency variable, such that the combined momentum-frequency integral is convergent at low frequencies and momenta. Whether a solution of Eq.~\eqref{eq: spingap equation} with $n_0 > 0$ and $m_s = 0$ exists, depends then on the size of the integral.
Carrying out the frequency integral, we obtain
\begin{equation} \label{eq: I def}
 {\cal I} = \int \frac{d\omega}{2\pi} \int_\qb
 \frac{2}{Z^\perp \omega^2 + J_{\alpha\beta}^\perp q_\alpha q_\beta} =
 \int_\qb \frac{1}{Z^\perp \omega_\qb} \, ,
\end{equation}
where $\omega_\qb = \sqrt{J_{\alpha\beta}^\perp q_\alpha q_\beta/Z^\perp}$.
The remaining momentum integral needs to be cut off at large momenta. We choose a cutoff scale $\Lambda_{\rm uv}$ such that the spin wave energy is restricted to values
$\omega_\qb < c_s^\perp \Lambda_{\rm uv}$, where $c_s^\perp = \sqrt{J^\perp/Z^\perp}$ with
$J^\perp = [{\rm det}(J_{\alpha\beta}^\perp)]^{1/2}$ is an average spin wave velocity.
With this choice, the momentum integral yields the particularly simple result
\begin{equation}
 {\cal I} = \frac{c_s^\perp \Lambda_{\rm uv}}{2\pi J^\perp} =
 \frac{\Lambda_{\rm uv}}{2\pi \sqrt{Z^\perp J^\perp}}  \, .
\end{equation}
Hence, Eq.~\eqref{eq: spingap equation} has a symmetry-broken solution with $m_s=0$ and $n_0 > 0$ if and only if
\begin{equation}
 Z^\perp J^\perp > \left(\frac{\Lambda_{\rm uv}}{2\pi}\right)^2 \, .
\end{equation}
This criterion depends on the rather arbitrary ultraviolet cutoff $\Lambda_{\rm uv}$. However, whatever value is chosen, since $J^\perp \to 0$ in the low doping limit $p \to 0$, one might conclude that ultimately $Z^\perp J^\perp$ will fall below the critical value $[\Lambda_{\rm uv}/(2\pi)]^2$, so that the ground state seems necessarily quantum disordered close to half-filling. This conclusion is however premature, since the momentum range in which the order parameter fluctuations can be described by the non-linear sigma model is shrinking to zero upon approaching half-filling. An analysis beyond the limitations of the sigma model will follow in the next section.


\subsection{Beyond the non-linear sigma model}

\begin{figure}[tb]
\centering
 \includegraphics[width=0.95\linewidth]{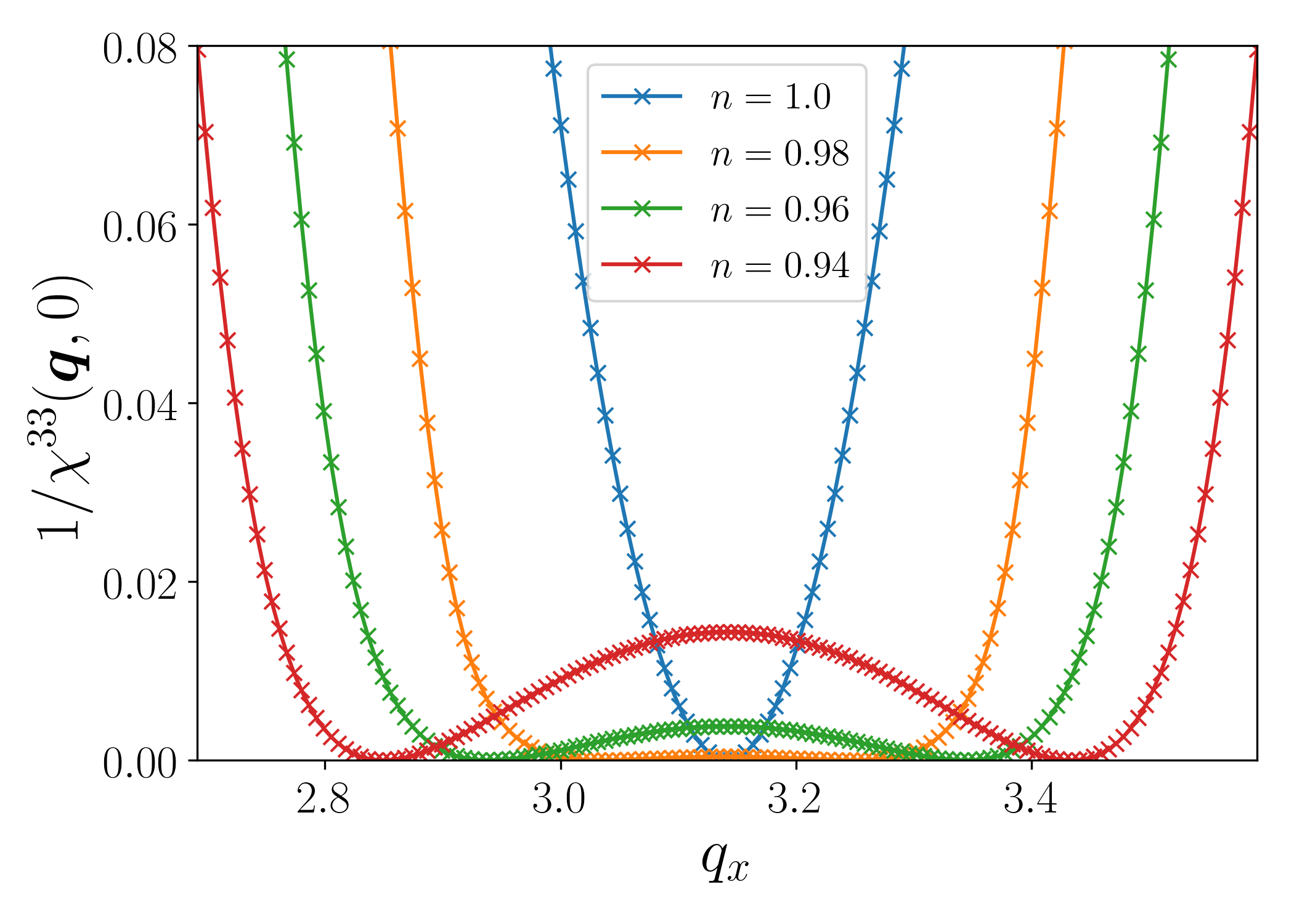}
 \includegraphics[width=\linewidth]{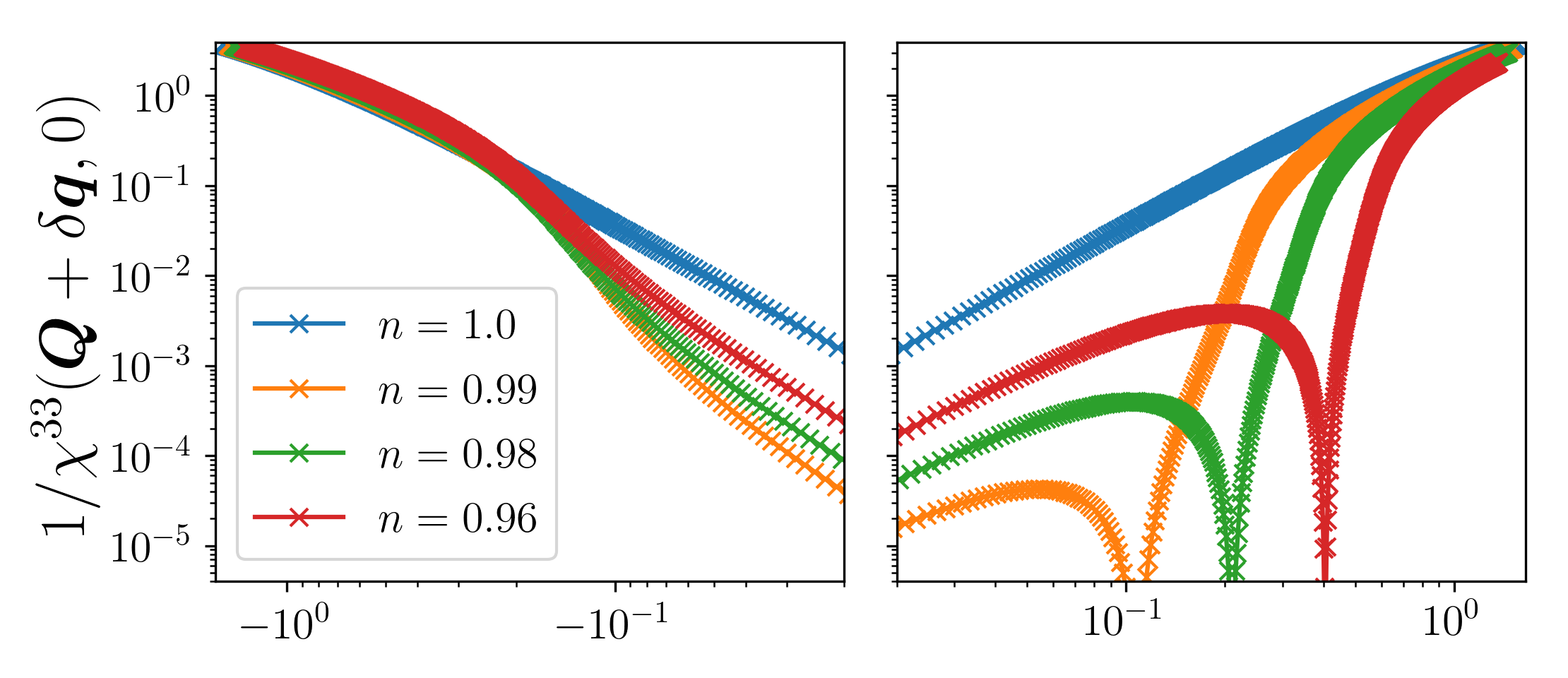}
 \caption{Top: Static inverse out-of-plane susceptibility $[\chi^{33}(\qb,0)]^{-1}$
 with $\qb = (q_x,\pi)$ as a function of $q_x$ for several densities at and slightly
 below half-filling.
 Bottom: Double logarithmic plot of $[\chi^{33}(\qb,0)]^{-1}$ with
 $\qb = \QQ + (\delta q_x,0)$ as a function of $\delta q_x$ for the same densities
 as in the top panel (left: $\delta q_x < 0$, right: $\delta q_x > 0$).}
 \label{fig: chi33_q}
\end{figure}

The description of order parameter fluctuations by the non-linear sigma model is based on an expansion of the spin susceptibility around the Goldstone poles to leading (quadratic) order in momentum and frequency. The coefficients of the leading quadratic momentum and frequency dependences are the stiffnesses. In Sec.~\ref{sec: II} we have shown that the spatial out-of-plane stiffness $J_{\alpha\beta}^\perp$ vanishes upon approaching half-filling from the hole-doped regime. We will now show that the momentum range in which the out-of-plane susceptibility is well described by the leading quadratic behavior, also shrinks to zero for $n \to 1^-$.

In Fig.~\ref{fig: chi33_q} we plot the inverse out-of-plane susceptibility
$[\chi^{33}(\qb,0)]^{-1}$ at zero frequency and $q_y = \pi$ as a function of $q_x$. The top panel shows a global view on a linear scale, and the double-logarithmic plot in the bottom panel highlights the regime near one of the Goldstone poles described by the leading quadratic momentum dependence.
At half-filling the momentum dependence of the inverse susceptibility is quadratic in a wide momentum range. Below half filling instead, the leading quadratic momentum dependence is limited to a much smaller momentum range, which shrinks to zero upon approaching half-filling.
It is clear that the momentum range supporting the leading quadratic momentum dependence near one of the Goldstone poles is limited by the vicinity of the other Goldstone pole (at which $[\chi^{33}(\qb,0]^{-1}$ also vanishes, see in particular the right panel at the bottom of the figure). The two poles, situated at $\QQ = (\pi-2\pi\eta,\pi)$ and $-\QQ = (-\pi+2\pi\eta,-\pi) \equiv (\pi+2\pi\eta,\pi)$ coalesce to $(\pi,\pi)$ as $\eta \to 0$. Hence, the momentum range in which a quadratic expansion is valid, necessarily vanishes with the incommensurability $\eta$, and thus with the doping $p$.

\begin{figure}[t]
\centering
 \includegraphics[width=\linewidth]{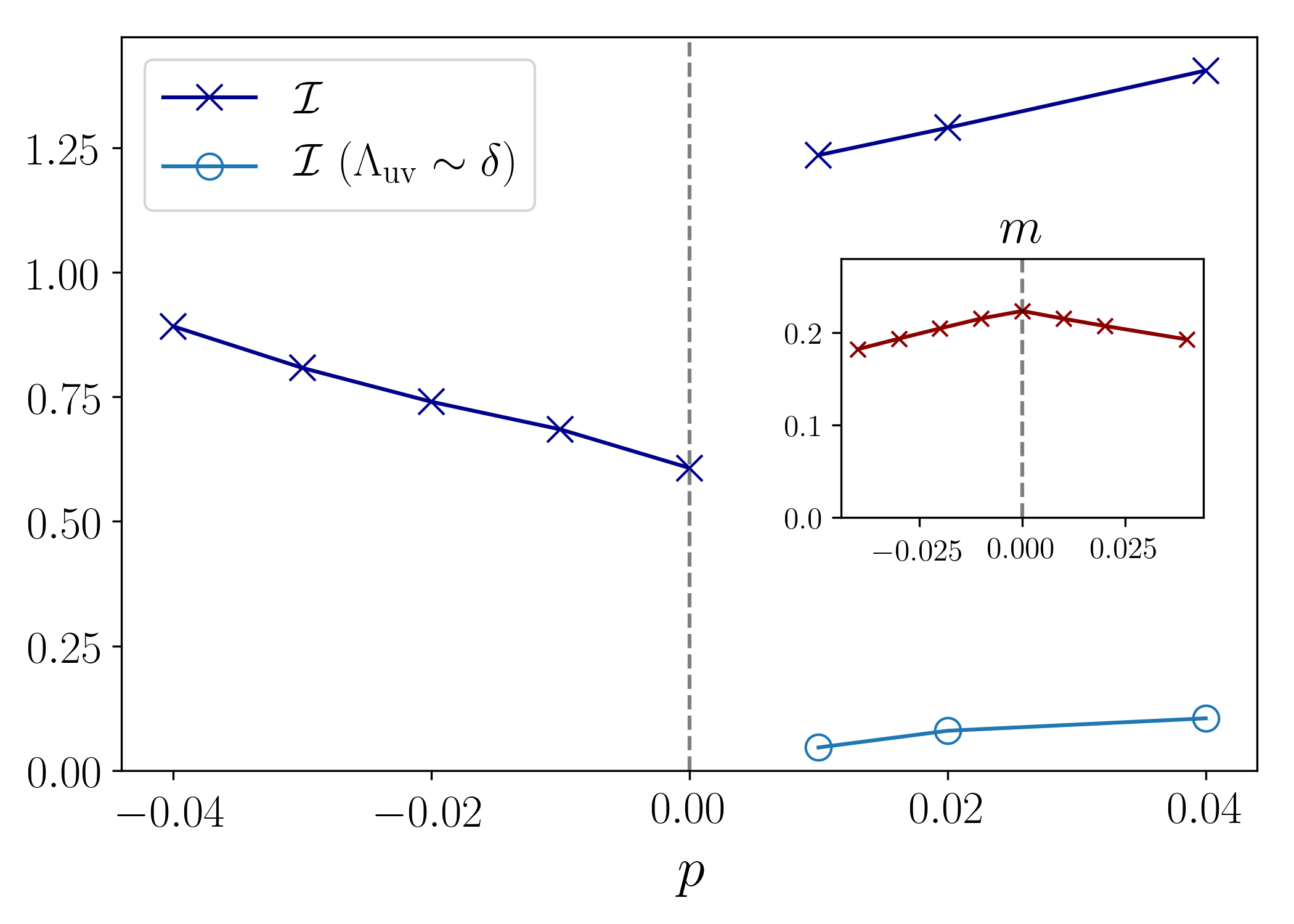}
 \caption{Momentum and frequency integral $\cal I$ over $m^{-2} \chi^{33}$ for various choices of hole and electron doping, along with the contributions from a small momentum region around one of the Goldstone poles (light blue curve), in which a quadratic expansion is applicable. The inset shows the magnetization $m$ as a function of doping.}
 \label{fig: tadpole}
\end{figure}

The integrand in Eq.~\eqref{eq: I def} corresponds to the quantity $m^{-2} \chi^{33}(\qb,i\omega)$, expanded to leading (quadratic) order around the two Goldstone poles at $\pm\QQ$, see Eq.~\eqref{eq: Goldstone33}. We have just shown that the expansion has a shrinking regime of validity in momentum space as we approach half-filling from the hole-doped regime. To better assess the stability of the magnetically ordered ground state in this regime, we now analyze the integral
\begin{equation}
\label{eq: tadpole}
 {\cal I} = \int \frac{d\omega}{2\pi} \int_\qb \frac{1}{m^2} \, \chi^{33}(\qb,i\omega)
\end{equation}
without expanding the integrand in momentum and frequency.
Note that $m^2 {\cal I}$ is a fluctuation correction to the mean-field order parameter amplitude square $m^2$, from out-of-plane spin fluctuations, which could also be derived from a perturbation expansion around the mean-field solution.
We cut off the momentum integral at a fairly large cutoff, $|\qb-(\pi,\pi)| \leq \Lambda_{\rm uv} = 1.2$, such that both Goldstone poles are situated deep inside the integration region for all considered doping values.

In Fig.~\ref{fig: tadpole} we show numerical results for the integral $\cal I$ for various choices of the doping at and near half-filling. We also show the size of the contributions from momenta close to one of the Goldstone poles, by imposing a small doping dependent cutoff $|\qb-\QQ| \leq \Lambda_{\rm uv}(p)$ with $\Lambda_{\rm uv}(p)$ proportional to $p$.
Inspite of the out-of-plane stiffness vanishing for vanishing hole doping $p \to 0^+$, the integral $\cal I$ remains finite in this limit. However, the numerical data indicate a discontinuity of $\cal I$ at half-filling on the hole doped side, that is, already a small hole doping leads to a large increase of $\cal I$ compared to its value at half-filling.

Exactly at half-filling, reliable numerical data for the magnetization of the two-dimensional Hubbard model are available \cite{Qin2016}. They show that the magnetic order parameter is reduced compared to mean-field calculations, but it remains finite. Due to the continuity of the fluctuation integral $\cal I$ upon electron doping, we can conclude that the ground state of the lightly electron doped Hubbard model remains magnetically ordered. The transition to a quantum disordered state can occur only at some finite critical doping value in the electron doped regime. By contrast, the discontinuous strong increase of $\cal I$ on the hole doped side indicates that the magnetic order could be destroyed by quantum fluctuations already for arbitrarily small hole doping.

We conclude this section by comparing our results with those by Kharkov {\em et al.}~\cite{Karkhov2018a}, who estimated an integral of the form integral~\eqref{eq: tadpole} near a N\'eel to spiral transition in frustrated magnets. In that work, the susceptibility was expanded as
\begin{equation}
\label{eq: chi Oleg}
    \chi^{33}(\bs{q},i\omega) \simeq \frac{m^2}{Z\omega^2 + \delta - a q^2 + b q^4} \ ,
\end{equation}
where $\bs{q}$ is the momentum relative to the N\'eel vector $(\pi,\pi)$.
This form captures both Goldstone poles at $\mathbf{Q}$ and $-\mathbf{Q}$, and, when these two modes are merging, the coefficients $a$ and $\delta$ necessarily tend to zero. Substituting this ansatz into the integral Eq.~\eqref{eq: tadpole}, one obtains
${\cal I} \sim b^{-1/2} \log\left(\Lambda_{\rm uv}/\sqrt{a}\right)$.
This term diverges logarithmically as $a \to 0$, provided $b$ remains finite, leading to a quantum disordered ground state near the (mean-field) N\'eel to spiral transition \cite{Karkhov2018a}.
However, in our case, a numerical analysis showed that the coefficient $b$ diverges as $p \to 0$, suppressing the logarithmic divergence, so that the integral remains finite.


\section{Conclusion} \label{sec: IV}

We have analyzed the doping dependence of the spin stiffnesses in the lightly doped two-dimensional Hubbard model. The results allowed us to address the open problem concerning the stability of magnetic order with respect to quantum fluctuations in the ground state.

The spin stiffnesses were obtained from an RPA calculation of the spin susceptibility in a magnetically ordered mean-field state. Choosing a moderate Hubbard interaction $U=2.35t$ and a sizable next-to-nearest neighbor hopping amplitude $t'=-0.2t$, the mean-field theory yields a N\'eel ordered ground state at and above half-filling, and spiral order with a wave vector of the form $\QQ = (\pi - 2\pi\eta,\pi)$ below. A functional renormalization group improved renormalized mean field theory \cite{Wang2014} with a Hubbard interaction $U = 4t$ yields roughly the same results \cite{Bonetti2022gauge}.
In the N\'eel state there are two Goldstone modes characterized by the same stiffness $J$, while the spiral state exhibits two Goldstone modes associated with out-of-plane fluctuations and a stiffness $J^\perp$, and another with in-plane fluctuations and a distinct stiffness $J^\Box$.

The asymmetry of the phase diagram between the electron doped ($n>1$) and the hole doped ($n<1$) regime is exceeded by an even stronger asymmetry of the doping dependence of the stiffnesses. Upon electron doping, the N\'eel stiffness $J$ decreases smoothly, and not very steeply. By contrast, upon hole-doping both $J^\perp$ and $J^\Box$ drop discontinuously at infinitesimal doping. The out-of-plane stiffness even drops to zero, and small finite values are recovered only at finite hole-doping $p=1-n$. We have derived several analytic formulae for the ordering wave vector and the spin stiffnesses in the lightly doped spiral regime. In particular, we have shown that the incommensurability $\eta$ is proportional to the doping $p$, with a prefactor depending exclusively on the spin stiffness at half-filling. The discontinuity of $J^\Box$ is fully determined by the effective masses of the hole-pockets, and we have proven analytically that $J^\perp$ indeed drops to zero.

We have also analyzed the doping dependence of the stiffness for a weaker interaction strength $U=2.2t$, where the ground state remains N\'eel ordered for small hole doping, before spiral order sets in via a weak first order transition at a critical doping strength $p_c$. In that case the stiffness also exhibits a pronounced drop at half-filling, and it vanishes upon approaching $p_c$ from below. On the spiral side of the transition the out-of-plane stiffness increases (from zero) continuously, while the in-plane stiffness jumps to a sizable finite value.

At finite temperatures the spin stiffnesses always vary continuously as a function of doping, and they vanish at the N\'eel-spiral transition. This agrees qualitatively with the doping dependence of the stiffnesses obtained at strong coupling in a recent dynamical mean-field calculation \cite{Goremykin2024}.

Based on our results for the density dependence of the spin stiffnesses, we could address the issue of the stability of magnetic order with respect to quantum fluctuations in the ground state of the two-dimensional Hubbard model. At half-filling (only), the ground state magnetization can be computed reliably by quantum Monte Carlo methods, and a finite magnetization has been obtained \cite{Qin2016}. The smooth dependence of the spatial and temporal spin stiffnesses as a function of electron doping therefore implies that magnetic order persists in some finite electron doping range above half-filling. By contrast, the drastic and discontinuous drop of the stiffnesses upon hole doping suggests a destruction of the magnetic order on the hole doped side of the phase diagram. However, the long wavelength expansion of the order parameter susceptibilities underlying the description of fluctuations by a non-linear sigma model breaks down in the lightly hole-doped regime. Our analysis of fluctuation corrections to the mean-field magnetization beyond the long wavelength expansion confirms their strong enhancement upon hole doping. Hence, the ground state of the lightly hole-doped Hubbard model is probably quantum disordered, at least for moderate interactions and in presence of a sizable next-to-nearest neighbor hopping.

In the context of SU(2) gauge theories of the pseudogap phase \cite{Bonetti2022gauge, Scheurer2018, Sachdev2019}, these results imply that the moderately interacting two-dimensional Hubbard model probably exhibits a robust pseudogap with a finite spin gap even for $T \to 0$ in the lightly hole doped regime, but not for light electron doping. This pronounced electron-hole asymmetry is in line with the experimental facts in the cuprates, where such a robust pseudogap is observed only in case of hole doping.

Our analysis can be extended to strong coupling, replacing the Hartree-Fock approximation by the dynamical mean-field theory \cite{Georges1996}. Preliminary results suggest that our qualitative findings on the electron-hole asymmetry of the spin stiffnesses, and the stability of the mean-field order with respect to quantum fluctuations, remain true at strong coupling.


\section*{Acknowledgements}

We are very greatful to A.~Chubukov, P.~Forni, K.~Fra\-boulet, A.~Georges, H.~M\"uller-Groeling, S.~Sachdev, T.~Sch\"afer, R.~Scholle, and O.~Sushkov for valuable discussions. P.M.B. acknowledges support by the German National Academy of Sciences Leopoldina through Grant No.~LPDS 2023-06, the Gordon and Betty Moore Foundation’s EPiQS Initiative Grant GBMF8683, and the U.S. National Science Foundation grant No.~DMR-2245246. 




\bibliography{main.bib}

\end{document}